\documentclass{sig-alternate}
\usepackage{graphicx}
\usepackage{tabularx}
\usepackage{balance}  % for  \balance command ON LAST PAGE  (only there!)
\usepackage{verbatim}
\usepackage{graphicx}
\usepackage[utf8]{inputenc}
\usepackage[export]{adjustbox}
\usepackage{listings}
\usepackage{wrapfig}
\usepackage[ruled,lined,linesnumbered]{algorithm2e} %option: linesnumbered, ruled
\usepackage{latexsym}
\usepackage{url}
\usepackage{caption} 
\usepackage{color}
\usepackage{multirow} 
\usepackage[tight,TABTOPCAP]{subfigure}
\usepackage{listings}
\usepackage{amsmath}
\usepackage{xfrac}
\usepackage{soul}
\usepackage{makecell}
\usepackage{enumitem}
\setlist{nolistsep,topsep=0pt,parsep=0pt,partopsep=0pt,leftmargin=12pt}

\DeclareMathOperator*{\argmin}{\arg\!\min}
\newcommand{\naive}{na\"{\i}ve}
\newcommand{\Naive}{Na\"{\i}ve}

\newcommand{\yeounoh}[1]{\textcolor{red}{YEOUNOH: #1}}

\newcommand{\michael}[1]{\textcolor[rgb]{1,0.4,0}{MICHAEL: #1}}

\newtheorem{definition}{Definition}

\makeatletter
\g@addto@macro\normalsize{%
  \setlength\abovedisplayskip{1pt}
  \setlength\belowdisplayskip{2pt}
  \setlength\abovedisplayshortskip{1pt}
  \setlength\belowdisplayshortskip{1pt}
}
\makeatother

\begin{document}

\numberofauthors{4}
\author{
\alignauthor Yeounoh Chung\\
	\normalsize{Brown University}
	\normalsize{yeounoh\_chung@brown.edu}
\alignauthor Michael Lind Mortensen\\
	\normalsize{Aarhus University}
	\normalsize{illio@cs.au.dk}
\alignauthor Carsten Binnig\\
	\normalsize{Brown University}
	\normalsize{carsten\_binnig@brown.edu}
\alignauthor Tim Kraska\\
	\normalsize{Brown University}
	\normalsize{tim\_kraska@brown.edu}
}

\title{Estimating the Impact of Unknown Unknowns\linebreak on Aggregate Query Results}

\newenvironment{packed_item}{
\begin{list}{$\bullet$}{
  \setlength{\topsep}{0pt}
  \setlength{\itemsep}{-5pt}
  \setlength{\parskip}{0pt}
  \setlength{\labelwidth}{15pt}
  \setlength{\leftmargin}{10pt}
  \setlength{\itemindent}{0pt}}
}{\end{list}}

\newenvironment{packed_item2}{
\begin{list}{}{
  \setlength{\itemsep}{-1.5pt}
  \setlength{\parskip}{1pt}
  \setlength{\labelwidth}{15 pt}
  \setlength{\leftmargin}{8pt}
  \setlength{\itemindent}{-8pt}}
}{\end{list}}

\maketitle
%\vspace*{-90pt}
\begin{abstract}
It is common practice for data scientists to acquire and integrate disparate data sources to achieve higher quality results. 
But even with a perfectly cleaned and merged data set, two fundamental questions remain: 
(1) is the integrated data set complete and 
(2) what is the impact of any unknown (i.e., unobserved) data on query results?

In this work, we develop and analyze techniques to estimate the impact of the unknown data (a.k.a., {\em unknown unknowns}) on simple aggregate queries. 
The key idea is that the overlap between different data sources enables us to estimate the number and values of the missing data items. 
Our main techniques are parameter-free and do not assume prior knowledge about the distribution. 
Through a series of experiments, we show that estimating the impact of {\em unknown unknowns} is invaluable to better assess the results of aggregate queries over integrated data sources. 
\end{abstract}

\section{Introduction}
\label{sec:intro}
In the past few years, the number of data sources has increased exponentially because of the ease of publishing data on the web, the proliferation of data-sharing platforms (e.g., Google Fusion Table \cite{FusionTables} or Freebase \cite{freebaseurl}), and the adoption of open data access policies, both in science and government.
The success of crowdsourcing \cite{WWWCrowdsourcing,crowddb,qurk,qurkdemo,scoop,crowdclouds,crowdsearch,HaasError} provides another virtually unlimited source of information.
This deluge of data has enabled data scientists, both in commercial enterprises and in academia, to acquire and integrate data from multiple data sources, achieving higher quality results than ever before.
It is therefore not surprising that industry and academia alike have developed highly sophisticated systems and tools to assist data scientists in the process of data integration \cite{data-integration-uncertainty-survey}. 
However, even with a perfectly cleaned and integrated data set, two fundamental questions remain: (1) do the data sources cover the complete data set of interest and (2) what is the impact of any unknown (i.e., unobserved) data on query results?

\vspace*{-4pt}
\subsection{Unknown Data}
\vspace*{-1pt}
In this work, we develop techniques to estimate the impact of the unknown data on aggregate queries of the form \texttt{SELECT AGGREGATE(attr) FROM table WHERE predicate}.

\begin{figure}[!t]
	%\vspace*{-75pt}
 \centering
 \includegraphics[height=1.4in,trim=5pt 10pt 5pt 5pt, clip=false]{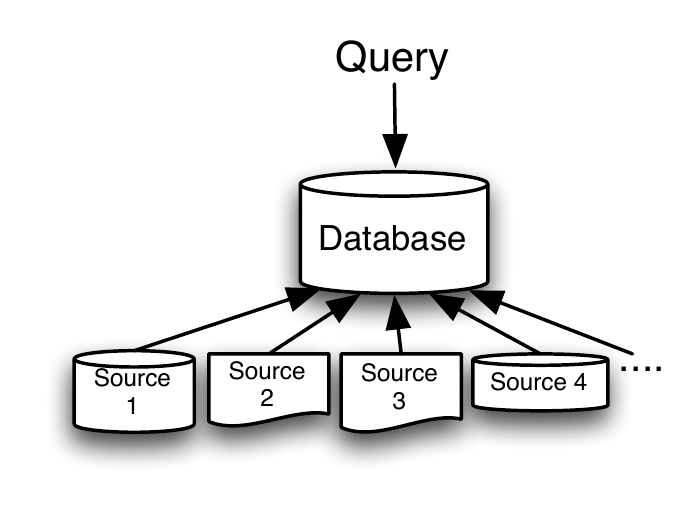}
  \vspace{-5pt}
 \caption{Simple data integration scenario where multiple data sources overlap but are not necessarily complete.}
 \vspace{4pt}
 \label{fig:integration_arch}
\end{figure}

We assume a simple data integration scenario, as depicted in Figure~\ref{fig:integration_arch}. 
Several domain-related data sources are integrated into one database, preserving the lineage information for each data item or record. 
Naturally, these data sources overlap with each other, but even when put together they might not be complete. 
For example, all data sources in Figure~\ref{fig:integration_arch} might list U.S. tech companies but some smaller companies might not be mentioned in any of the sources. 
This data integration scenario applies to a wide range of use cases ranging from crowdsourcing  (where every crowd-worker can be considered a single data source \cite{crowddb}) to data extraction from web pages.

Estimating the impact of the unknown data (data items that are not observed in any data source) is particularly difficult as we neither know how many unique data items are missing and their values; thus, we deal with {\bf unknown unknowns}.
This characteristic distinguishes our work from what is generally known as {\em missing data}, or {\em known unknowns}, estimation in Statistics \cite{allison2012handling,rubin1976inference,rahm2000data}, which tries to estimate the value of unknown (missing) attributes for known records.
At a first glance, it may seem impossible to estimate the impact of {\em unknown unknowns}; however, for a large class of data integration scenarios, the analysis of overlap of multiple data sources makes it feasible.

\vspace*{-4pt}
\subsection{A Running Example} \label{sec:running_example}
\begin{figure}[t]
 \centering
 \includegraphics[width=2.4in]{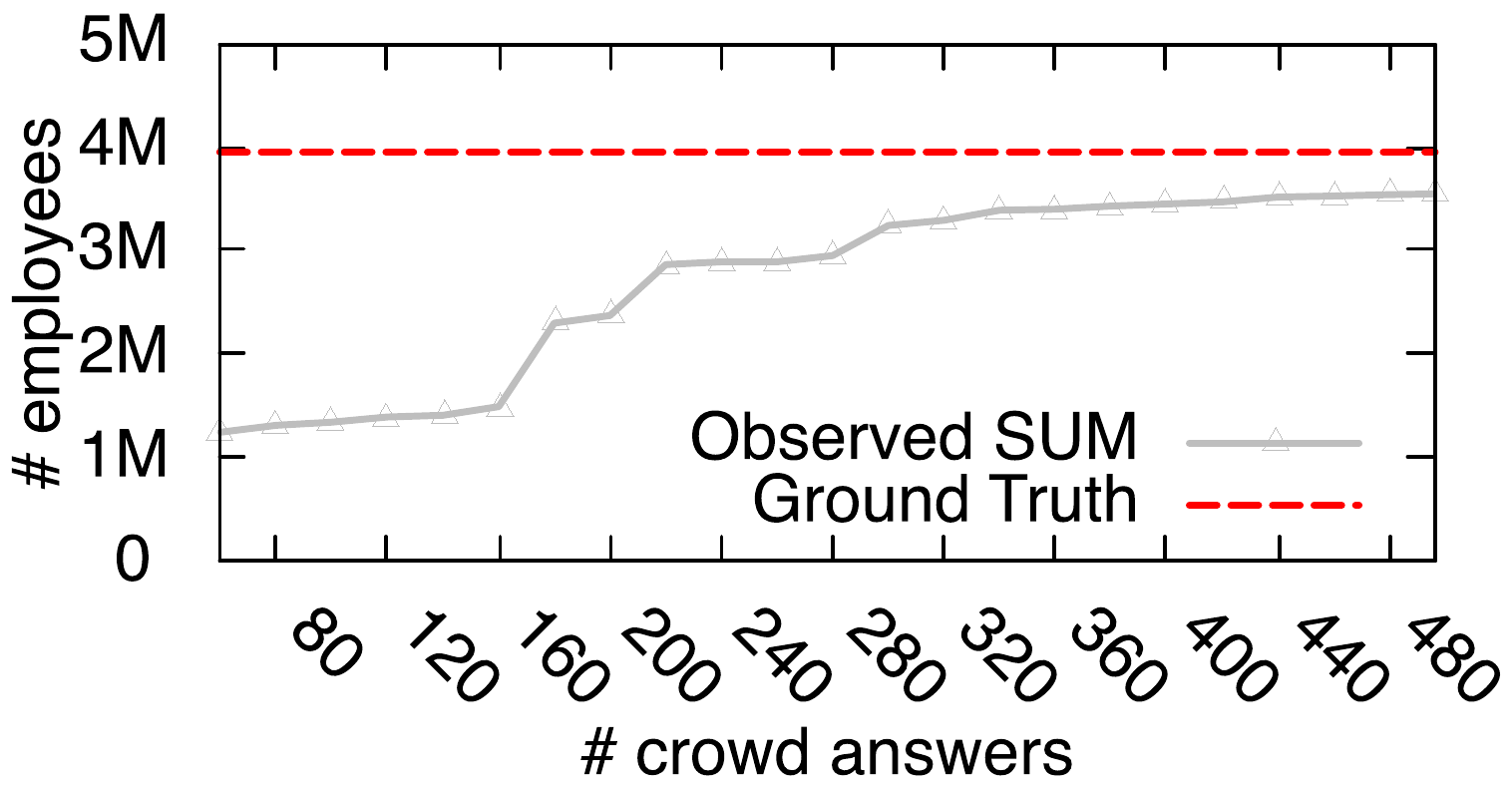}
 \caption{Employees in the U.S. tech sector}
 \label{fig:intro}
% \vspace{-17pt}
\end{figure}

\vspace*{-1pt}
To demonstrate the impact of {\em unknown unknowns}, we pose a simple aggregate query to calculate the number of all employes in the U.S. tech industry, \texttt{SELECT SUM(employees) FROM us\_tech\_companies}, over a crowdsourced data set. 
We used techniques from \cite{crowddb} to design the crowdsourcing tasks on Amazon Mechanical Turk (AMT) to collect employee numbers from U.S. tech companies.\footnote{More precisely, we only asked for companies with a presence in Silicon Valley, as we found it provides more accurate results (see also Section~\ref{sec:eval}).}
The data was manually cleaned before processing (e.g., entity resolution, removal of partial answers). 
Figure~\ref{fig:intro} shows the result. 

The red line represents the ground truth (i.e., the total number of employees in the U.S. tech sector)  for the query \cite{real_data_empl}, whereas the grey line shows the result of the observed SUM query over time with the increasing number of received crowd-answers. 
The gap between the observed and the ground truth is due to the impact of the {\em unknown unknowns}, which gets smaller at a diminishing rate as more crowd-answers arrive.

While the experiment was conducted in the context of crowdsourcing, the same behavior can be observed with other types of data sources, such as web pages. 
For instance, suppose a user searches the Internet to create a list of all solar energy companies in the U.S. 
The first few web pages will provide the greatest benefit (i.e., more new solar companies), while after a dozen web pages the benefit of adding another web page diminishes as the likelihood of duplicates increases. 
The rate of increasing overlap of data sources is indicative of the completeness of the data set. 

\vspace*{-4pt}
\subsection{A Na\"ive Solution}
\vspace*{-1pt}
The same type of diminishing effect is also known as the \emph{Species Accumulation Curve} in Ecology \cite{ugland2003species}, where the rate of new species discovered decreases with increasing cumulative effort to search.
Measuring species richness (i.e., counting species) is critical in many ecological studies. 
Plotting a \emph{Species Accumulation Curve} provides a way to estimate the number of additional species to be discovered. 

These species estimation techniques lay the foundation for estimating the impact of unknown unknowns on aggregate query results. 
A \naive{} solution for the SUM query from Section \ref{sec:running_example}  would be to first estimate the number of {\em unknown data} items using  species estimation techniques \cite{gettingitall} and then use {\em mean substitution} to estimate their value \cite{osborne2012best}.
This assumes that the missing items have on average the same attribute value as the observed (known) data items. 

The \naive{} approach has a couple of drawbacks. 
First, species estimation has very strict requirements on how data is collected.
Almost every data integration scenario violates these requirements, causing the estimator to significantly over/underestimate the number of missing data items. 

Second, it ignores the fact that the attribute values of the missing items may be correlated to the likelihood of observing certain data items.
For example, large tech companies like Google with many employees are often more well known and thus, appear more often in data sources than smaller start-ups, creating a biased data set. 
This is problematic as it also biases the mean and with it the estimate.

In the statistics literature, this second problem is referred to as {\em Missing Not At Random} (MNAR)  \cite{osborne2012best,rubin1976inference}, where the missingness of a data item depends on its value. 
There are many statistical inference techniques dealing with MNAR \cite{allison2012handling,dempster1977maximum,d2000estimating,allison2012handling,yuan2010multiple,rahm2000data}, but nearly all the techniques require at least partial knowledge of the record. 
For example, in the case of surveys, people with a high salary might be more reluctant to report their salary but have no problem stating their home address or how many children they have. 
Existing MNAR techniques use the reported values (e.g., the address) to infer the missing attributes. 
Unfortunately, this is not possible in the case of {\em unknown unknowns}, as we miss the entire record. 

\vspace{-4pt}
\subsection{Contributions}
\vspace*{-1pt}
This work is a first step towards developing techniques to estimate the impact of the {\em unknown unknowns} on query results.
Our focus is on simple aggregate queries, especially {\em SUM}-aggregates, but we also touch upon other aggregations like {\em COUNT}, {\em AVG}, {\em MIN}, and {\em MAX}. 
We design techniques that can deal with the peculiarities of the data integration scenarios discussed before, such as uneven contributions from different sources (bias of data sources). 

In this work, we use crowdsourced data sets because they are easier to collect, but the techniques are general and apply to almost all data integration scenarios that combine overlapping data sources.
While we do not argue that the proposed techniques can predict black-swan-like data items (i.e., extremely rare data items), we will show that our techniques can provide useful estimates under more ``normal'' circumstances, which we will define more formally. 
For instance, in the example of Figure~\ref{fig:intro} we can get an almost perfect estimate of the impact of the {\em unknown unknowns}  after only 350 crowd-answers.
In addition, by building upon recent work on the Good-Turing estimator \cite{mcallester2000convergence}, we are able to provide an upper bound for our estimates under easy to understand conditions.
In summary we make the following contributions:

\begin{itemize}
\item We formalize the problem of estimating the impact of {\em unknown unknowns} on query results and describe why existing techniques for species estimation and missing data estimation are not sufficient.
\item We develop techniques to estimate the impact of the {\em unknown unknowns} on aggregate query results.
\item We derive a first upper bound for {\em SUM}-aggregate queries.
\item We examine the effectiveness of our techniques via experiments using both real and synthetitc data sets.
\end{itemize}

In the following, we first formalize our problem statement (Section~\ref{sec:problem}), presents techniques to estimate the impact of {\em unknown unknowns} for sum-queries (Section~\ref{sec:sum})  and propose an upper bound estimate (Section~\ref{sec:upper_bound}). Section~\ref{sec:other_query} extends these techniques then to other aggregate functions and in Section~\ref{sec:eval} we evaluate our techniques, followed by related work and conclusion.

%The paper is organized as follows: In Section~\ref{sec:problem} we formalize our data integration scenario and the meaning of {\em unknown unknowns}. In Section~\ref{sec:sum} we present techniques to estimate the impact of {\em unknown unknowns} on {\em SUM-}aggregate queries, while our upper bound is described in Section~\ref{sec:upper_bound}. Section~\ref{sec:other_query} extends these techniques to other aggregate functions and in Section~\ref{sec:eval} we evaluate our techniques using real-world and simulated data sets. Section~\ref{sec:related} discusses related works, and we conclude in Section~\ref{sec:concl}. 

\section{The Impact of Unknown\\ Unknowns}
\label{sec:problem}

In this Section, we define {\em unknown unknowns}, explain how data integration over multiple sources can be regarded as a sampling process and formally define our estimation goal. 
For convenience Appendix~\ref{appendix:symbols} contains a symbol-table.

For the purpose of this work, we treat data cleaning (e.g., entity resolution, data fusion, etc.) as an orthogonal problem. Any data cleaning techniques \cite{allison2012handling, FlorescuDI, flexiblequery, completeness, Neiling00dataintegration, 754917} can be applied to our problem without altering the problem context. While data quality can influence the estimation quality, studying it goes beyond the scope of this paper \cite{gettingitall}. 
We assume that after a proper data cleaning process we have one instance per observed entity and know exactly how many times the entity was observed across multiple data sources.

\subsection{Unknown Unknowns}
We assume that queries are of the form \texttt{SELECT AGGREGATE(attr) FROM table WHERE predicate}, that $table$ only contains records about a single entity class (e.g., companies) and that a record in $table$ corresponds to exactly one real-world entity (e.g., IBM).
Thus, in the remainder of the paper we use record, entity and data item interchangeably.

\vspace*{-5pt}
\begin{definition}\label{def:unknown_data} (Unknown Unknowns) 
	Let $\Omega$ be the universe of unknown size of all valid unique entities $r$ for a given entity class and $attr_A(r)$ be the value of attribute $A$ of $r$. 
	Then the ground truth $D \subseteq \Omega$ is defined as a set of entities that satisfy the predicate, i.e., $D=\{r\in \Omega ~|~predicate(r)\}$, where its size $N=|D|$ is not known. 
	Let $S$ be a sample with replacement from $D$ and $c$ be the number of unique entities in $S$.
	Unknown unknowns $U$ refers to any unobserved entity $r$ that exists in $D$ but not in $S$: $U = D - S$ with size $N - c$.
\end{definition}
\vspace*{-5pt}

\begin{figure}[t!]
 \centering
 \includegraphics[width=2.5in]{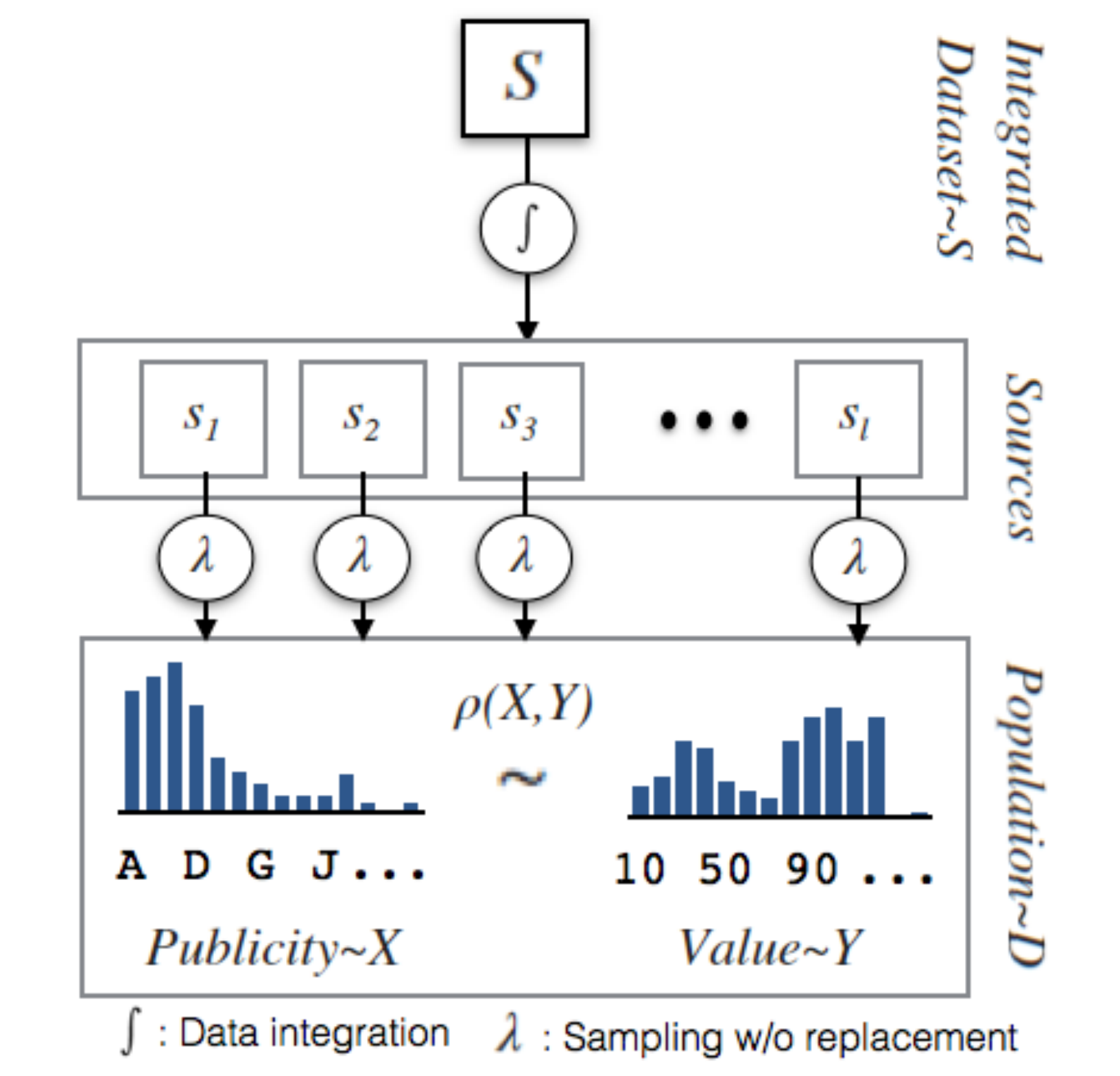}
 \caption{A sampling process for the integrated database. 
% Publicity distribution~X and value distribution~Y can be potentially correlated.
}%; however, two different items $x_1$ and $x_2$ can have the same value $y$, but very different publicity (i.e., likelihood to be sampled) $X(x_1) \neq X(x_2)$.}
 \label{fig:sampling}
  \vspace{2pt}%\vspace{-17pt}
\end{figure}

For our running example, $\Omega$ would be the universe of all companies in the world, $D$ all tech companies in the US and $\sum_{r \in D}attr_{empl}(r)$ be the true number of U.S. tech sector employees.   
$S$ would be a sample with duplicates and {\em unknown unknowns} would be every company which is not in  $S$.
 
What we aim to achieve is a good estimate of the ground truth: \texttt{SELECT AGGREGATE(attr) FROM D}, when we only have $S$. 
Note, that we drop the $predicate$ from the query, since every item in $D$ already has to fulfill the $predicate$. 
In this work, we assume that we neither know all entities in $D$ nor its size (i.e., {\em open world} assumption). This distinguishes our problem from the problem of {\em missing data} \cite{osborne2012best,rubin1976inference,diveshtutorial}, which refers to incomplete data or missing attribute values. 

%In the next subsection, we explain how data integration can be regarded as a sampling process.

%However, we may use the techniques to deal with missing data (e.g., mean substitution, imputation) \cite{allison2012handling} to estimate the values of unknown data.

%In the next sub-section we explain, why the process of data integration can be regarded as a sample process. 

%\begin{definition}\label{def:unknown_data} (Unknown Data) Let $S$ be a sample with $c$ unique data items, sampled from an underlying population $D$. The number of unique data items in $D$ or $N=|D|$ is not known ($c \leq N$). Any unobserved item $i$ that exists in $D$, but not in $S$, is defined as unknown data.
%\end{definition}

%If the population size $N$ is not known, there is a possibility of unknown data. Note that this is different from the traditional notion of missing data, which refers to incomplete data or missing attribute values. However, we may use the techniques to deal with missing data (e.g., mean substitution, imputation) \cite{allison2012handling} to estimate the values of unknown data.

\vspace{-4pt}
\subsection{Data Integration As Sampling Process}
\label{sec:problem:integration}\vspace{-1pt}
Data integration refers to the process of combining different data sources under a common schema \cite{FlorescuDI}.
For the purpose of this work, we assume that data sources are independent samples (e.g., data source are not copies from each other and instead are independently created), and we model the data integration process as a multi-stage sampling process as shown in Figure~\ref{fig:sampling}.

We assume $l$ data sources $s_1 ... s_l$, each sampling $n_j=|s_j|$ data items from the ground truth $D$ (e.g., the complete set of tech companies in the US with their respective number of employees), {\bf without replacement},  as a data source typically only mentions a data item once. 
We assume further that every data item $d_i \in D$ has a {\em publicity} likelihood $p_i$ of being sampled, following some distribution $X$. 
Likewise, the attribute values (e.g., the number of employees) have a certain likelihood to appear in the ground truth, referred to as {\em value likelihood}, again following some distribution $Y$. 
These two distributions are possibly correlated making the {\em publicity-value correlation} bigger or smaller than 0: $\rho \neq 0$.

The  data sources $l$ are then integrated into a single integrated data set $S$ of size $n_{S} = \sum_{j=1}^l n_j$. 
Although each source samples without replacement from $N=|D|$ different classes (i.e., unique data item), $S$ contains duplicates because every data source is sampling from the same underlying truth $D$.
If $l$ is sufficiently large, {\bf $S$ approximates a sample with replacement} from $D$, which is the reason why species estimation techniques work in the first place (we analyze the effects of smaller $l$ in Section~\ref{sec:sum:mc} and~\ref{sec:eval}).
The number of unique data items $c$ in $S$ is likely to be smaller than $N$. 
In contrast, the end-user only sees a view of $S$, referred to as the integrated database $K$ (for {\em Known} data), which contains only one entity per unique entity in $S$. 
%Note, that we require entity resolution as part of this process to determine how often we observed a specific data item in $S$. 
%\carsten{Do we really need $S$ and $K$??? Or can we just say that $S$ has only one tuple per real world entity and a count of how often it appeared in the underlying sources.}
%\tim{That it something we discussed often. Both is possible.}

This data integration model covers a large class of use cases from web integration to crowdsourcing.
In the latter case, each crowd worker can be regarded as a separate data source $s_j$ as it is known that workers also sample without replacement from $D$ \cite{gettingitall}.
While extremely powerful, there are scenarios where this sampling model does not apply.
Most importantly, data sources are not always independent \cite{diveshairline}.
Furthermore, the number of data sources $l$ has to be large enough to have sufficient overlap between the sources (see Section~\ref{sec:eval}). 
If any of these assumptions are violated, then only low-quality estimations are possible.

\vspace{-3pt}
\subsection{Problem Statement}\vspace{-1pt}
We are interested in estimating the impact of {\em unknown unknowns} ($U$) to adjust aggregate query results.

\vspace*{-5pt}
\begin{definition}\label{def:unknown_impact} (The Impact of Unknown Unknowns) 
Given an integrated database $K$, the impact of unknown unknowns is defined as the difference between the current answer $\phi_{K}$ of the aggregate query over the database $K$ and the answer over the ground-truth $\phi_{D}$:
\begin{equation}
 \Delta =  \phi_{D} - \phi_{K}
\end{equation}
\end{definition}
 
Our goal is to estimate the answer on the ground-truth by estimating $\Delta$ based on $S$:
\begin{equation}
 \hat{\phi}_{D} = \phi_{K} + \hat{\Delta}(S)
 \label{eqn:main}
\end{equation}

%\vspace*{-5pt}

Note that this definition works for all common aggregates including \emph{MIN} and \emph{MAX}, where $ \hat{\Delta}$ would be the positive or negative adjustment to the observed \emph{MIN}/\emph{MAX} value.

%Given an aggregate query, any unknown item $i$ that exists in $D$, but not in $S$, will offset the query result by some amount pertaining to its attribute value, from the ground truth query result over $D$.

%To estimate the {\em impact of unknown}, in specific, we need to estimate the number of {\em unknown data} (e.g., US Tech companies) and their values (e.g., number of employees). We can use the {\em impact of unknown} $\phi_{adj}$ to adjust the query result $\phi_{S}$; the ground truth query result estimate $\hat{\phi_{D}}$ is defined as follows:
%\begin{equation}\label{eqn:naive}
%	\hat{\phi}_{D} = \phi_{S} + \phi_{adj}
%\end{equation}

\vspace*{-8pt}
\section{Sum Query} \label{sec:sum}
In this section, we focus on {\em SUM}-aggregates to illustrate our estimation techniques.  We first formalize the \naive{} estimator (Section~\ref{sec:sum:naive}), which was informally introduced in the introduction. 
We then develop the {\em frequency} estimator by making {\em \naive{}} estimator more robust to the {\em  publicity-value correlation} (Section~\ref{sec:sum:freq}). 
Afterwards, we describe the more sophisticated {\em bucket} estimator  (Section~\ref{sec:sum:bucket}).
%As our experiments will show the {\em bucket} estimator has the highest accuracy for data sets with a {\em publicity-value correlation}.
Finally, we develop a {\em Monte-Carlo} estimator which is better suited for a smaller set of data sources (Section~\ref{sec:sum:mc}).
%In Section~\ref{sec:other_query} we then show how the same techniques can be used for  other aggregates like {\em AVG} or {\em MIN/MAX}.

\vspace*{-4pt}
\subsection{Na{\"i}ve Estimator}\vspace*{-1pt}
\label{sec:sum:naive}
Estimating the impact of {\em unknown unknowns} for $SUM$ queries is equivalent to solving two sub-problems: (1) estimating how many unique data items are missing (i.e., the {\em unknown unknowns} count estimate), and (2) estimating the attribute values of the missing data items (i.e., the {\em unknown unknowns} value estimate). The {\em \naive{}} estimator uses the $Chao92$ \cite{chao92} species estimation technique to estimate the number of the missing data items, and \emph{mean substitution} \cite{osborne2012best} to estimate the values of them. 

Let $\phi_K = \sum_{r \in K}{attr(r)}$  be the current sum over the integrated database, then  we can more formally define our {\em \naive{}} estimator for the impact of {\em unknown unknowns}  as:
\begin{equation}\label{eqn:adj_chao}
\Delta_{naive} = \underbrace{\frac{\phi_K}{c}}_{\text{Value estimate}}\cdot \underbrace{(\hat{N}-c)}_{\text{Count estimate}}
\end{equation}
$\hat{N}$ is the estimate of the number of unique data items in the ground truth $D$, and $c$ is the number of unique entities in our integrated database $K$ (thus, $\hat{N}-c$ is our estimate of the number of the unknown data items). 
$\phi_K/c$ is the average attribute value of all unique entities in our database $K$. 

%Note that the correction to $\phi_{K}$ using \naive{} estimator is simply the product of the expected number of missing unknown data items and the average {\em attribute} value of the observed data items. 

%In the following, we discuss $Chao92$ estimator to sketch the general ideas behind species estimation techniques and show how $Chao92$ incorporates skewness of sample in its estimation.

%In the following, we only discuss the widely-used $Chao92$ \cite{chao92} estimator to sketch the general ideas behind species estimation techniques. Furthermore, as $Chao92$ is more robust to {\em publicity-value correlation} as it incorporates sample skew in its estimation.
%We use $Chao92$ throughout this work because it incorporates data skew in its estimation.

\vspace*{-4pt}
\subsubsection{Chao92 estimator}\label{sec:sum:chao92}\vspace*{-1pt}
%All species estimation techniques  have one thing in common: they require a sample {\em with} replacement from $D$.
%Yet our model specifies that all $l$ data sources sample without replacement from $D$.
%The reason why we can still use these techniques is, that if the number of data sources is large enough, $S$ approximates a sample with replacement over $D$.
%However, if the number of data sources is small, the estimation quality will suffer. 
%In the extreme case of one data source, no estimation is possible (most techniques will estimate an infinite number of unknown data items). 
%Similar if one data source is significantly larger than others, $\exists j:  n_j \gg n_i  \forall i \neq j $, it can have a similar effect and significantly over-estimate the number of unknown data items.

%In the following, we assume that the number of data sources is large enough and contribute roughly the same amount of data items (in our experiments we found that 5 data sources are often sufficient), whereas in Section~\ref{sec:sum:mc} we develop a new estimator, which explicitly considers the number of data sources and their sizes $n_j$. 
%Of course, we will also study the effect of the number of data sources in our experiment section. 

Throughout the paper, we use the popular $Chao92$ estimator. Many species estimation techniques exist \cite{bunge_review93,chao_review05}, but we choose $Chao92$ since it is more robust to a skewed publicity distribution. 
The \emph{Chao92} estimator uses \emph{sample coverage} to predict $\hat{N}$. 
The sample coverage $C$ is defined as the sum of the probabilities $p_i$ of the observed classes. Since the true distribution ${p_1 ... p_N}$ is unknown, we estimate $C$ using the Good-Turing estimator \cite{turing53}: 
\begin{equation}
\hat{C} = 1 - f_1/n
\label{eqn:coverage}
\end{equation}
The $f$-statistics, e.g., $f_1$, represent the frequencies of observed data items in the sample, where $f_j$ is the number of data items with exactly $j$ occurrences in the sample. 
$f_1$ is referred as {\em singletons}, $f_2$ {\em doubletons}, and $f_0$ as the {\em missing data} \cite{bo78}. 
Sample coverage measures the ratio between the number of singletons ($f_1$) and the sample size ($n$). 
This ratio changes with the amount of duplicates in the sample.
The high-level idea is the more duplicates that exist in our sample $S$ compared to the number of singletons $f_1$, the more complete the sample is (i.e., higher sample coverage). 

In addition, the \emph{Chao92} estimator explicitly incorporates the skewness of the underlying distribution using \emph{coefficient of variance} ($CV$) $\gamma$, a metric that is used to describe the dispersion in a probability distribution \cite{chao92}. 
%The $CV$ is defined as the standard deviation divided by the mean. 
A higher $CV$ indicates a higher variability among the $p_i$ values, while a $CV=0$ indicates that each item is equally likely (i.e., the items follow a uniform distribution). 
 
Given the publicity ($p_1 \cdots p_N$) that describe the probability of the $i$-th class being sampled from $D$, with mean $\bar{p} = \sum_i p_i/N = 1/N$,  $CV$ can be expressed as follows:
\begin{equation} \label{eqn:gamma}
\gamma = \left[\sum_i(p_i - \bar{p})^2/N\right]^{1/2} \left. \right/ \bar{p}
\end{equation}
However, since $p_i$ is not available for all data items, $CV$ has to be estimated using the $f$-statistic: 
\begin{equation}
\label{eqn:cv_est}
\hat{\gamma}^2 = \max \left\{\frac{\frac{c}{\hat{C}} \sum_i{ i(i-1)f_i}}{n(n-1)} - 1 \, , \,0\right\}
\end{equation}
The final  \emph{Chao92} estimator for $\hat{N}_{Chao92}$ can then be formalated as: \vspace*{-8pt}
\begin{equation}
\hat{N}_{Chao92} = \frac{c}{\hat{C}} + \frac{n(1-\hat{C})}{\hat{C}}\cdot\hat{\gamma}^2
\label{eqn:chao92}
\end{equation}

\vspace*{-5pt}
\subsubsection{The Estimator}\vspace*{-1pt}
$\hat{N}_{Chao92}$ is our estimate for $N$, and comparing this to $c$ provides us with a means of evaluating the completeness of $S$. By substituting $\hat{N}_{Chao92}$ for $\hat{N}$, the final {\em \naive{}} estimator can be written as:
\begin{equation}
\begin{split}
\Delta_{naive} = \frac{\phi_{K}}{c}\cdot (\hat{N}_{Chao92}-c) 
= \frac{\phi _K \cdot f_1 \cdot \left(c+\hat{\gamma }^2 n\right)}{c \cdot \left(n-f_1\right)}  
\label{eqn:naive}
\end{split}
\end{equation}

Note, that the {\em \naive{}} estimator does not consider any {\em publi-\\city-value correlation} and thus tends to over- or under-estimate the ground truth. 
%In our running US tech company example, the estimator will overestimate the missing data as bigger companies are more well-known, yielding a biased (higher) mean. 

\vspace*{-2pt}
\subsection{Frequency Estimator}
\label{sec:sum:freq}
We developed a simple variation of the {\em \naive{}} estimator, which makes direct use of the frequency statistics to improve estimation quality. 
All coverage-based species estimation methods give special attention to the singletons $f_1$; the data items observed exactly once. 
The idea is that those items, in relation to the sample size $n$, give a clue about how well the complete population is covered. 
A ratio of $f_1 / n$ close to $1$ means that almost every sample is unique, indicating that many items might still be missing. 
Conversely, a ratio close to $0$ indicates all unique values have been observed several times, decreasing the likelihood of any unknown data. 
We use a similar reasoning to improve our value estimation. 
The key idea is that singletons are the best indicator of missing data items, and that their average value might be a better representation of the values of the missing items. 
Let $\phi_{f_1}$ be the sum of all singletons, $\sum_{r \in singletons} attr(r)$  and $\hat{N}_{Chao92}$ again be the $Chao92$ count estimate. Then the estimator can be defined as:   
\begin{equation}\label{eqn:adj_f1}
\begin{split}
\Delta_{freq} &= \frac{\phi_{f_1}}{f_1}\cdot (\hat{N}_{Chao92}-c) 
= \frac{\phi _{f_1} \left(c+\hat{\gamma }^2 n\right)}{n-f_1}
\end{split}
\end{equation}

While this estimator still does not directly consider the {\em publicity-value correlation}, it is more robust against popular high-impact data items (i.e., data items with extreme {\em attribute} values).
For example, in our running employee example, big companies that are highly visible like Google or IBM can significantly impact the known value estimate $\phi_{K}/c$.
However, through using  the average value of the singletons, $\phi_{f_1}/f_1$, it is reasonable to assume that those companies will not stay as singletons very long in any sample and thus will not impact the average value for the {\em unknown unknowns}. 
This estimator is surprisingly simple and becomes even simpler if we assume $\hat{\gamma}^2=0$:  
\begin{equation}\label{eqn:freqsimple}
\Delta_{freq}=\frac{\phi_{f_1}\cdot c}{n-f_1}
\end{equation}

Note, that $\hat{\gamma}^2=0$ makes it a Good-Turing estimate, which also converges to the ground truth even for skewed publicity values; it might just take a bit longer \cite{chao92}.
While $\Delta_{freq}$ is not the best estimator  (see Section~\ref{sec:eval}) the simplicity  makes it still useful to quickly test if an aggregate query result might be impacted by any {\em unknown unknowns}. 

\vspace*{-4pt}
\subsection{Bucket Estimator}\vspace*{-1pt}
\label{sec:sum:bucket}
The problem with the previous two estimators is that they do not directly consider a correlation between publicity and attribute values. 
We designed the {\em bucket} as a  first estimator designed for {\em unknown unknowns} with {\em publicity-value correlation}. 
The idea of the estimator is to divide the attribute value range into smaller sub-ranges called buckets, and treat each bucket as a separate data set.
We can then estimate the {\em impact of unknown unknowns} per bucket (e.g., large, medium, or small companies) and aggregate them to the overall effect: \vspace*{-7pt}
\begin{equation}\label{eqn:bucket}
\Delta_{bucket} = \sum_{i}{\Delta(b_i)}
\end{equation}

Here $\Delta_{b_i}$ refers to the estimate per bucket and both the {\em frequency} or {\em \naive{}} estimator could be used.
%We can actually use the \naive{} or {\em frequency} estimator per bucket. 
Using buckets has two effects: First, it provides a more detailed estimate on what types of companies are missing and related to that, second, the value variance per bucket decreases, making the estimate less prune to outliers (e.g., items with extreme low and high values can be ``contained'' in separate buckets).

The challenge with the {\em bucket} estimator is to determine the right size for each bucket. 
If the bucket size is too small, the bucket contains almost no data items.
In an extreme case of having a single data item per bucket, no count or proper value estimation is possible.
If the bucket size is too big, then the {\em publicity-value correlation} can still bias the estimate. 
In fact, the case with a single bucket is equivalent to using just the {\em \naive{}} or {\em frequency} estimator.
In the following we describe two bucketing strategies. 
%We now describe how buckets are defined both manually and automatically, to cope with any degrees of {\em publicity-value correlation}.

\vspace*{-4pt}
\subsubsection{Static Bucket}\label{sec:sum:bucket:static}\vspace*{-1pt}
An easy way to define buckets is to divide the observed value range into a fixed $n_b$ number of buckets of size $w_i$:
\begin{equation}\label{eqn:bucket_static}
w_i = \frac{(a_{max}-a_{min})}{n_{b}}
\end{equation}

where $a_{min}$ ($a_{max}$) refers to the min (max) observed attribute value.
Afterwards we apply $\Delta_{naive}$ per bucket. 
It is important to note that the estimate goes to infinite with buckets which only contain singletons due to division-by-zero ($n-f_1=0$, see equation~\ref{eqn:naive}), which can significantly increasing the error of the estimate for very small buckets. 
%On the other hand, using fewer buckets is not much different than using the {\em \naive{}} or {\em frequency} estimators.

Unfortunately,  the optimal number of buckets varies depending on the underlying publicity distribution (see Appendix~\ref{appendix:static}). When the publicity distribution is more skewed and correlated to attribute values, some static buckets may contain too few data items, whereas others contain more than enough. The true publicity distribution is not known and we cannot predetermine the right number (or size) of static buckets. To this end, we found that static buckets based estimation is of little practical value.

%\carsten{Isn't it a better argument to say that the static bucket technique disregards the distribution of data items in $S$ and thus might produce empty buckets or buckets with too little elements (for non-uniform distributions)? Moreover, for uniform distributions static bucketing is not helpful either!}

%In general, having too few buckets will suffer from data skew (for count estimation) or even sample bias (for value estimation), whereas having too many buckets would leave each bucket with insufficient data to perform any estimations.
\vspace*{-4pt}
\subsubsection{Dynamic Bucket} \label{sum:bucket:dynamic}\vspace*{-1pt}
To overcome the previously mentioned issues, we developed several alternative statistical approaches to determine the optimal bucket boundaries over time.
The most notable are our uses of the error estimate/upper bounds from Section~\ref{sec:upper_bound} and of treating $f_1$ as a random variable  (see also Section~\ref{sec:sum:other}).
Surprisingly, we achieve the best performance across all our real-world use cases and simulations using a rather simple conservative approach, referred to as $\Delta_{Dynamic}$.

The core idea behind our  dynamic strategy $\Delta_{Dynamic}$ is to sort the {\em attribute} values of $S$ and then recursively split the range into smaller buckets only if it minimizes the estimated impact of {\em unknown unknowns}, i.e., the absolute $\Delta$ value. 
Intuitively, this is controversial since either under- or overestimation could be better for different use cases. 
However, there is a more fundamental reason behind this strategy.

{\bf The Foundation:}
Whenever we split a data set into buckets, each bucket contains less data than before the split, and the chance of an estimation error increases due to the {\em law of large numbers} (i.e., the less data the higher the potential variance) \cite{mcallester2000convergence, n_e_cv}.
To illustrate this, we consider the simplest case of a uniform {\em publicity} distribution ($\hat{\gamma}=0$) and an even bucket split. 
In this case, we can show that the $Chao92$  estimate  for $\hat{N}$ is bigger or equal to the $Chao92$  $\hat{N}$ before the split: \vspace*{-7pt}
\begin{equation} \label{eqn:bucket_ineq}
\begin{split}
\hat{N}_{Chao92} & = \frac{c}{1-f_1/n} = \overbrace{\frac{n\cdot c}{n-f_1}}^{\text{Before split}}\\
 & \leq \underbrace{\frac{n_{b1} \cdot c_{b1}}{n_{b1} - f_{1_{b1}}} + 
 \frac{n_{b2} \cdot c_{b2}}{n_{b2} - f_{1_{b2}}}}_{\text{After split}}
\end{split}
\end{equation}

When we split the data exactly into halves, it follows that $c_{b1} = c_{b2} = c/2$ (i.e., we split in regard to the unique values). 
With a uniform publicity distribution, every item is equally likely, and therefore we can assume that both buckets contain roughly the same amount of data after the split: $n_{b1} = n_{b2} \approx n / 2$.
However, in contrast to $n$ and $c$, the number of singletons ($f_1$) can vary significantly between the buckets.
In fact, we know that the estimators only stabilize if every item was observed several times \cite{chao92} and as a consequence  $n$ has to be significantly larger than $c$ and $c$ significantly larger than $f_1$ ($n \gg c \gg f_1$). 
Therefore, the variance of $f_1$ is relatively higher than the one of $n$ or $c$ between the buckets and if we split, there is a higher chance that we unevenly distribute the $f_1$ among the buckets.
%whereas $n$ and $c$ are more evenly distributed. 

\begin{algorithm}[tb]
\small
\SetAlgoLined
\SetKwInOut{Input}{Input}\SetKwInOut{Output}{Output}
\Input{ $S$ }
\Output{List of buckets}
\BlankLine
    $b_0 = (minValue(S), maxValue(S))$ \tcc*[r]{init bucket $b_0$}
    $todo = [b_0]$\tcc*[r]{list with $b_0$}
    $\delta_{min} = abs(\Delta(b_0))$  \tcc*[r]{$\Delta$ estimate over $b_0$}
    $bkts = []$ \tcc*[r]{final bucket list}
    \While{$!todo.empty$}{
        $b = todo.pop$ \tcc*[r]{remove first element}
        $\delta_{tmp} = \delta_{min} - abs(\Delta(b))$\;
        $tmp = (null,null)$ \tcc*[r]{Empty pair}
        \For{$unique\: r \in b$}{
            $(t_1,t_2) = split(b,r.value)$ \; 
            \If{$\delta_{min}>\delta_{tmp} + abs(\Delta(t_1)) + abs(\Delta(t_2))$}{
               $\delta_{min} = \delta_{tmp} + abs(\Delta(t_1)) + abs(\Delta(t_2))$\;
               $tmp = (t_1, t_2);$
            }
        }
        \uIf{$tmp \neq (null,null)$}{
               $todo.add(t_1, t_2);$
        }\Else{
            $bkts.add(b)$\;
        }
    }
    \BlankLine
    return $bkts$\; 
    \caption{Dynamic bucket generation} 
    \label{alg:dynamic_bkt}
\end{algorithm}

%is that $c$ and $n$ can be assumed to be relatively large even after the split, whereas they might only be a few singletons $f_1$. 
%In fact,  the estimators stabilize if and only if every item was observed several times and as a consequence  $n$ has to be significantly larger than $c$ and $c$ significantly larger than $f_1$. 
%\begin{equation}
%n \gg c \gg f_1
%\nonumber
%\end{equation}
%Thus, the variance of $f_1$s is relatively higher than the variance of $n$ or $c$ between the buckets. 
%If we now split the buckets, there is a high chance that we unevenly distribute the $f_1$ among the buckets while $n$ and $c$ are roughly equally split between them. 
To model the uneven distribution of $f_1$ we introduce another parameter $\alpha \in [0,1]$ and set $f_{1_{b1}} = \alpha \cdot f_1$ and $f_{1_{b2}} =  ( 1- \alpha ) \cdot f_1$. As a result the inequality in equation~\ref{eqn:bucket_ineq} becomes:
\begin{equation} \label{eqn:bucket_ineq_2}
\underbrace{\frac{n\cdot c}{n-f_1}}_{\text{Before split}} \leq
\underbrace{\frac{\frac{n}{2}\cdot\frac{c}{2}}{\frac{n}{2}-\alpha \cdot f_1} + \frac{\frac{n}{2}\cdot\frac{c}{2}}{\frac{n}{2}-(1-\alpha)\cdot f_1}}_{\text{After split}}
\end{equation}

Appendix~\ref{appendix:increase_count} shows that the right hand side of the above inequality has its global minimum at $\alpha=0.5$, which evaluates to $nc/(n-f_1)$ ($\hat{N}$ before split), and that the inequality always holds.
Thus, it can be seen that splitting a data set into buckets not only potentially increases the error, but it does so in a monotonic way. 
%It follows that whenever the  estimate of $\Delta$ increases after a split, it is likely due to the increasing error in $\hat{N}$ (an overestimation of the {\em unknown unknowns} count estimate).

%DO NOT TOUCH THIS SECTION BEFORE TALING TO TIM
Yet, this does not mean that the {\em sum} estimate $\Delta$ always increases as well. 
Especially with a {\em publicity-value correlation}, the overall estimate of $\Delta$ over all buckets can still decrease as the average {\em attribute} values per bucket differ. 
This is in-line with our original  motivation to use buckets, as we wanted to get a more detailed {\em unknown unknowns} estimate (e.g., how many small companies vs. large companies are missing). 
Bringing these two observations together, we can assume for many real-world use cases that  whenever our estimate of the impact of unknown unknowns $\Delta$ increases after a split, it has a significant chance of being caused by the increasing error in $\hat{N}$, whereas when it decreases it potentially improves the estimate due to the more detailed unknown estimate. 
While it does not always have to be the case (e.g., if the publicity-value correlation is negative) it is still an indicator for many real-world use cases (see Section~\ref{sec:eval}).
Based on the observations, we have devised the conservative bucket splitting strategy: only split the bucket if the overall estimate for $\Delta$ is minimized.

{\bf The Algorithm:}
Algorithm~\ref{alg:dynamic_bkt} shows the final algorithm.
First we add a bucket which covers the complete value range of $S$ to the $todo$ list (line 2) and calculate the current $\Delta$ over S (line 3). 
Note that we take the absolute values of all estimates ($\Delta$) to underestimate the impact of {\em unknown unknowns} even for the case of having negative attribute values (e.g., net losses of companies).
Afterwards, we check recursively if we can split the bucket to minimize $\Delta$ until no further ``underestimation'' is possible (line 5-21).

We therefore remove the first bucket from the $todo$ list (line 6) and calculate the $\Delta$ over $S$ without the impact of this bucket $b$ (line 7). 
Note, that during the first iteration  $\delta_{tmp}$ will be 0.
Afterwards, for every unique record in $b$, we split the current bucket $b$ into two temporary buckets $t_1$ and $t_2$ based on the record's attribute value (line 10).
If the resulting estimate using this split is bigger than any previously observed minimums (line 11), we set the new minimum to this value (line 12) and temporally store the new buckets (line 13). 
When the for-loop of line 9-15 finishes and if at least one new bucket was found (line 16), $tmp$ will contain the new  split point, which minimizes $\delta$ for the bucket, and $\delta_{min}$ the new minimum value of $\delta$.
Those buckets are then added to the $todo$ list (line 17) to be checked, if splitting them again would further lower the estimate.
On the other hand, if $tmp$ is empty, the algorithm wasn't able to further split the bucket and the current bucket without any additional splits is added to the final bucket list (line 19). 
If no buckets are left in the $todo$ list, the algorithm terminates and $bkts$ contains the final list of buckets.

\vspace*{-4pt}
\subsection{Monte-Carlo Estimator}\vspace*{-1pt}
\label{sec:sum:mc}
As our experiments show, the previous estimator actually performs very well (see Section~\ref{sec:eval}). 
However, what it does not consider is the effect of uneven contributions from data sources (i.e., one data source contains much more data than another) and the peculiarities of the sampling process itself. 
The $Chao92$ species estimation, like almost all other estimators, assumes sampling {\em with} replacement, whereas our data sources sample {\em without} replacement from the underlying ground truth.
The reason why the  $Chao92$ still works is, that with a  reasonably high number of data sources the integrated data source $S$ approximates a sample with replacement \cite{gettingitall}.
However, with either a small number of data sources or uneven contributions from sources (i.e., some sources are significantly bigger than others), $S$ diverges significantly from a sample done with replacement, resulting in significant over- or under-estimation. 
In the case of crowd-sourcing, the workers which provide significantly more data items than other workers, are referred to as {\em streakers} \cite{gettingitall}.

To address these issues, we  present a {\em Monte Carlo}-based (MC)  estimator for $\hat{N}$.
The idea  is that we simulate the sampling process to find the best distribution with its population size $N$, which best explains the observed sample including how many items $s_j$  every data source $j$ contributes. 
More formally, given  $(s_1,...,s_l)$ what we seek is a set of parameters $\Theta$ (e.g., the distribution parameters) for the MC simulation,  which minimize some distance function $\Gamma$ between the observed data $S$ and the simulated data $Q_{\Theta}$:
\begin{equation}\label{eqn:mc}
\argmin_{\Theta}  \Gamma(S, Q_{\Theta} | l, [s_1,...,s_l])
\end{equation}

In the following we first describe the MC method for generating $Q_{\Theta}$ with given $\Theta$, the distance function $\Gamma$, and finally the search strategy to find the optimal parameter $\Theta$.

\vspace*{-3pt}
\subsubsection{Monte-Carlo Method} \vspace*{-1pt}
\label{sec:mc:method}
In contrast to the other estimators, the {\em Monte-Carlo} estimator requires an assumption about the shape of the underlying {\em publicity} distribution;
in this work, we use an exponential distribution for {\em publicity}, from which  data source $j$ samples $n_j$ data items.
Accordingly, the parameter $\Theta$ has two components: $\theta_{N}$ specifies the assumed number of data items, and $\theta_{\lambda}$ governs the shape (skew) of {\em publicity} distribution.
Note, that the assumption of the exponential distributions makes the MC method a parametric model. 
The goal of the MC simulation is to determine how well  $\theta_{N}$ and $\theta_{\lambda}$ help to explain the observed $S$.

Algorithm~\ref{alg:mc_method} shows our MC algorithm. 
First, we use an exponential distribution with skew $\theta_{\lambda}$ to sample {\em publicity} ($p_1 \cdots p_{\hat{N}}$) for $\theta_{\hat{N}}$ items (line 1). And then we initialize the distance to 0 (line 2).
Afterwards we repeat the following procedure $nbRuns$ times.  
For every data source (line 5) we sample $n_j$ data items  according to $E$, but also  without replacement (line 6). The sampled items are added to $Q$ to form a histogram (line 7) for the particular run. 
After simulating $l$ sources, $Q$ contains the simulated version of $S$.

To finally compare the simulated sample $Q$ with the observed sample $S$, we make use of the discrete KL-divergence metric \cite{lexa2004useful}.
However, this requires transforming $S$ and $Q$ into a frequency statistic and indexing them to ensure that the right items are compared with each other (line 9).

After the indexing we have two comparable frequency statistics for $S$ and the simulation: $F_S$ and $F_Q$.
However, $S$ might contain less than $\hat{N}$ unique data items, for which the KL-divergence is not defined. 
We therefore adjust $F_S$ and assign a small non-zero probability to the missing extra unique items (line 10).
Finally, the two frequency statistics can be compared using the standard KL-Divergence metric and added to the total distance (line 11) and after all the simulation runs the average distance is returned (line 13).
%\begin{equation}\label{eqn:kl_disc}
%D_{KL}(F_{S}, F_Q)=\sum_{i}F_{S}(i)\cdot\ln{\frac{F_{S}(i)}{F_{Q}(i)}}
%\end{equation}

\begin{algorithm}[tb]
\small
%\IncMargin{1em}
\SetAlgoLined
\SetKwInOut{Input}{Input}\SetKwInOut{Output}{Output}
%\Indm
\Input{ $\theta_{\hat{N}}$, $\theta_{\lambda}$, $S$, $[n_1,...,n_l]$, $nbRuns$}
\Output{Average distance}
%\Indp
\BlankLine
    $E = dist(\theta_{\hat{N}},\theta_{\lambda})$\tcc*[r]{publicity of $\hat{N}$ items}
    $\Gamma = 0.0$\tcc*[r]{default value}
    \BlankLine
    \For{$i = 1$ \KwTo $nbRuns$}{
        $Q = []$\tcc*[r]{simulated model}
        \For{$j = 1$ \KwTo $l$}{
            $s_i=sample(n_j,E)$\tcc*[r]{w/o repl}
            $Q.add(s_i)$\;
        }
        $(F_S, F_Q) = indexing(S,Q)$\;
        $F_S' = smooth(F_S, F_Q)$\;
        $\Gamma\mathrel{+}=klDiv(F_S',F_Q)$\tcc*[r]{KL-divergence}
    }
    \BlankLine
    return $\Gamma /nbRuns$\;
    \caption{Monte Carlo method}
    \label{alg:mc_method}

\end{algorithm}

\vspace*{-5pt}
\subsubsection{Search Strategy} \label{sec:mc:optimization}\vspace*{-1pt}
We can now simulate the observed sampling process leading to $S$, but we still need a way to find the optimal $\Theta$, which best explains the observed sample $S$. 
The difficulty is, that even though the KL-divergence cost function is convex, the integer variable $\hat{N}$ prevents us from using tractable optimization algorithms (e.g., gradient descent). 
Furthermore, the the distance function can be quite sensitive to small amounts of noise in $D$. 

We therefore make the estimator more robust by first performing a grid search for $\Theta$ (line 5-10). We vary $\theta_N$ between $c\leq \hat{N} \leq \hat{N}_{Chao92}$ with a step-size $(\hat{N}_{Chao92}-c)/10$ and $\theta_{\lambda}$ between $-0.4\leq \lambda \leq 0.4$ (i.e., almost no to heavy skew) with a step-size $0.1$ (line 2 and 3).
%Note that $\lambda$ values $-0.4$ and $0.4$ represent a heavily skewed publicity distribution, so we cover a wide range of skew levels. 
The step sizes are chosen to be small enough to efficiently model the convex curve, but large enough to be robust to any noise.
Afterwards, we fit a two-dimensional curve using least-squares curve fitting (line 11) and return the $\hat{N}_{MC}$ with the minimum $D_{KL}$ on the fitted curve as the final count estimate (line 11). 

Finally, to estimate the total difference, we use our {\em \naive{}} estimation technique with $\hat{N}_{MC}$. 
The estimate is more robust and over-estimates less than the original {\em \naive{}} estimator as our MC method always penalizes any unmatched unique items in $Q$. In other words, the MC estimator favors solutions where $\hat{N}$ is closer to the number of observed unique items $c$. 

\vspace*{-3pt}
\subsection{Other Estimators}\vspace{-1pt}
\label{sec:sum:other}
During the course of developing the above estimators, we explored various alternatives. 
For example, we experimented with alternative static bucket strategies (see also Appendix~\ref{appendix:static}).
Most importantly though, we noticed that many proposed techniques can actually be combined. 
For instance, we can use the {\em frequency} estimator, instead of the {\em \naive{}} estimator, with the {\em bucket} (i.e., {\em Dynamic Bucket} approach) estimator or the {\em Monte-Carlo} estimator.
More interestingly, we can also combine the {\em Monte-Carlo} estimator with the {\em bucket} estimator. 
%The {\em Monte-Carlo} method in Algorithm~\ref{alg:mc_method} uses an exponential distribution to model the  {\em publicity} distribution, which can take various forms. So by applying the {\em Monte-Carlo} method per bucket, which covers a smaller value range of possibly more uniform {\em publicity} (with non-zero {\em publicity-value correlation}) items, we can make our choice of exponential distribution (also an uniform distribution when $\lambda=0$) more suitable. 
However, as the {\em Monte-Carlo} estimator requires large sample sizes to be accurate, we found that it often decreases the estimation quality. Similarly, we found that the difference between the {\em \naive{}}  and {\em frequency} estimators does not help much for the {\em bucket} approach (see Appendix~\ref{appendix:other_estimators}). 
For the experiments we therefore focus on the original techniques rather than the various combinations and included the other results in the appendix. 

\begin{comment}
We can also employ different splitting conditions for {\em Dynamic Bucket} estimator. In essence, the Algorithm~\ref{alg:dynamic_bkt} splits bucket to prevent $Chao92$ from overestimating (i.e., failing with insufficient duplicates). An alternative is to split to improve unknown data value estimation (i.e., mean substitution) using the relationship between number of examples $n$, data skew $\gamma$ and margin of error $e$ in estimating the population mean \cite{n_e_cv}:
\begin{equation}\label{eqn:lynch_kim}
n = \frac{z^2 \cdot \gamma^2}{e^2}
\end{equation}
For instance, at least 3 data items with $\gamma \leq 0.05$ are needed to estimate the unknown statistic, $\phi$, within $5\%$ margin of error at $95\%$ confidence (i.e., $z=1.96$).
\end{comment}

%\DecMargin{1em}

%\DecMargin{1em}

\begin{algorithm}[tb]
\small
%\IncMargin{1em}
\SetAlgoLined
\SetKwInOut{Input}{Input}\SetKwInOut{Output}{Output}
%\Indm
\Input{$[s_1, ... s_l]$,$c$,$\hat{N_{Chao92}}$,$nbRuns$}
\Output{Estimated number of unique data items, $\hat{N}$}
%\Indp
\BlankLine
    $D_{KL}=[]$\tcc*[r]{KL-divergence}
    $n=sizes([s_1, ..., s_l])$\tcc*[r]{$[n_1,...,n_l]$}
    $\Theta_{\hat{N}} = [c:\frac{(\hat{N}_{Chao92}-c)}{10}:N_{Chao92}]$\;
    $\Theta_{\lambda} = [-0.4:0.1:0.4]$\;
    \BlankLine
    \For{$\theta_{\hat{N}} \in \Theta_{\hat{N}}$}{
        \For{$\theta_{\hat{N}} \in \Theta_{\lambda}$}{
            $\Gamma = monteCarlo(\theta_{\hat{N}},\theta_{\lambda},n,n_r)$\tcc*[r]{Alg 2}
            $D_{KL}.add(\Gamma)$\;
        }
    }
    \BlankLine
    $p = curveFit(\Theta_{\hat{N}},\Theta_\lambda,D_{KL},2)$\tcc*[r]{2-D curve fit}

    $[\hat{N},\lambda] = \underset{\lambda \in [-0.4,0.4], \hat{N} \in [c,\hat{N_{Chao92}}]}
                {\arg\min \{p(\hat{N},\lambda)\}}$\tcc*[r]{min on the curve}
    \BlankLine
    return $\hat{N}$\;
    \caption{Monte-Carlo based $\hat{N}$ estimation}
    \label{alg:mc_estimator}
\end{algorithm}
\vspace*{-4pt}
\section{Estimation Error Upper Bound} \label{sec:upper_bound}
In this section, we derive an estimation error upper bound, specifically, the worst case estimation error of the {\em \naive{}} estimator (Equation~\ref{eqn:adj_chao}). 
The same upper bound can easily be applied to each bucket in the {\em bucket} estimator, as well as the {\em Monte-Carlo} estimator. 

To estimate the impact of {\em unknown unknowns} on \emph{SUM} query results we multiply the  estimate for the number of unknown data with the estimate of the values. 
Hence, we define the worst case estimate as the product of the worst case unknown data count and the worst case value estimate. 

The $Chao92$ count estimation is based on {\em sample coverage} plus a correction for the skew $\hat{\gamma}>0$.
Recent work proposed a tight error bound of the Good-Turing estimator for the ground truth {\em unknown unknowns} distribution mass ($M_0$) \cite{mcallester2000convergence}: \vspace*{-7pt}
\begin{equation}\label{eqn:good_turing_bound}
M_0 \leq 
\frac{f_1}{n} + (2\sqrt{2}+\sqrt{3})\cdot\sqrt{\frac{\log{3/\epsilon}}{n}}
\end{equation}
 which holds with {\bf probability at least $1-\epsilon$} over the choice of the sample with $n=|S|$. The confidence parameter $\epsilon$ governs the tightness of this bound (we use $\epsilon=0.01$ for $99\%$ confidence). Based on equation~\ref{eqn:good_turing_bound}, we bound $Chao92$: 
\begin{equation}\label{eqn:upper_bound}
\begin{split}
\hat{N}_{Chao92} & = \frac{c}{\hat{C}} + \frac{n(1-\hat{C})}{\hat{C}}\cdot\hat{\gamma}^2\\
  & \approx \frac{c}{\hat{C}} = \frac{c}{1-M_0}\\
  & \leq 
\frac{c}{1-(\frac{f_1}{n} + (2\sqrt{2}+\sqrt{3})\cdot\sqrt{\frac{\log{\log{3/\delta}}}{n}})}
\end{split}
\end{equation}
Notice, that we can omit $\hat{\gamma}$ as it only makes the $Chao92$ converge faster, but does not influence the asymptotic estimate, which is based on the sample coverage. 

%Notice that we simplify $\hat{N}_{Chao92}$ by ignoring any data skew ($\hat{\gamma}$). In general, $Chao92$ converges faster by incorporating the data skew \cite{gettingitall} but by definition a bound o
%and eventually converges to $c/\hat{C}$ as we collect more data (i.e., $\hat{C}\rightarrow 1$) and the second term ($n(1-\hat{C})/\hat{C}\cdot\hat{\gamma}^2$) becomes negligible. 
 
%and ignore $\hat{\gamma}$ as it will still remain the upper bound\footnote{The proof of this is trivial as Chao92 only guarantees faster convergence if data skewed exist}

%However, if we would take advantage of $\hat{\gamma}$ we might be able to achieve a tighter upper bound: 
%(val-v_est)^2/(c-1); (c_sum/c + 3*std)*(C_hat-c)

As the distribution of the {\em mean substitution} ($\frac{\phi_K}{c}$) tend to a normal distribution ({\em Central Limit Theorem}), we define the worst case estimate of the ground truth {\em attribute} mean value ($\frac{\phi_D}{N}$) with the help of the sample standard deviation ($\sigma_K$): \vspace*{-7pt}
\begin{equation}
\frac{\phi_D}{N} \leq \frac{\phi_K}{c} + z\cdot\sigma_K
\end{equation}
Here $z$ controls the confidence of the bound, and we use $z=3$ based on the three-sigma rule of thumb \cite{three_sigma} to have nearly all  values with $99.95\%$ confidence lie below the upper bound.
The final upper bound is then the simple multiplication of  the two worst case estimators (we present the results in Section~\ref{sec:eval_other}): \vspace*{-7pt}
\begin{equation}
\begin{split}
  \Delta_{bound} & = \frac{(\frac{\phi_K}{c} + z\cdot\sigma_K) \cdot c}{1-(\frac{f_1}{n} + (2\sqrt{2}+\sqrt{3})\cdot\sqrt{\frac{\log{\log{3/\delta}}}{n}})}
\end{split}
\end{equation}

%\tim{You assume a certain confidence here, correct? For both the upper-bound for the count and the value? Should we not better show the general version, where the user can choose the confidence he wants} \yeounoh{$\delta$ is the confidence parameter for count, and z is now the confidence parameter. but, it is conventional to use 3.}
\vspace*{-8pt}
\section{Other Aggregate Queries}\label{sec:other_query}
In this section we describe how the same techniques for {\em SUM}-aggregates  can be applied to other aggregates for estimating the impact of the unknown unknowns. 

{\bf COUNT:} Estimating {\em COUNT} is easier than {\em SUM} as it only requires estimating the number of unknown data items, but not their values.
For instance, one could either directly use the $Chao92$ estimator or the techniques proposed in \cite{gettingitall}.
In addition, the {\em bucket} and {\em Monte-Carlo} approaches can be used simply by skipping the second step, i.e., not multiplying the estimated count with the value estimates.

{\bf AVG:} The simplest way to estimate the {\em AVG} with {\em unknown unknowns} is to use the {\em AVG} over the observed sample $S$ (i.e., the law of large numbers). This is reasonable because of the  law of large numbers. 
However,  $S$ might be biased due to a {\em publicity-value correlation} and need to be corrected. 
One way to deal with the bias is to use our {\em bucket} approach with a simple modification on how the $\Delta_b$ per bucket are aggregated (e.g., weighted average of averages by the number of unique data items ($\hat{N}_{Chao92}$) per bucket). 
%(averages can not simply be added up and divided).

{\bf MAX/MIN:} At a first glance, it seems impossible to estimate {\em MIN} or {\em MAX} in the presence of {\em unknown unknowns}. However, we can still do better than simply returning the observed extreme values by reporting when we believe that the observed minimum or maximum value is the true extreme values. This is already very helpful in many integration scenarios and easy to do with our {\em bucket} estimator. The strategy divides the observed value range of $S$ into consecutive sub-ranges (i.e., buckets); the number of {\em unknown unknowns} as well as their values are estimated per bucket. If the estimated {\em unknown unknowns} count in the highest (lowest) value range bucket is zero, then we say that we have observed the true maximum (minimum) value and only then report the highest (lowest) value.  

%However, as they identify unknown data and complete the database, the proposed techniques can also be used for other aggregate query processing (e.g., COUNT, AVERAGE, MIN and MAX) of a form \texttt{SELECT AGGREGATE(attribute) FROM relation WHERE predicate}.

%While we believe we can estimate the impact of more regular errors, we do not argue that we can devise methods which predict black swan events \cite{blackswanbook,blackswan}; events that are so irregular that they are almost impossible to predict (e.g., a solar energy company that nobody knows about but that has more revenue than the biggest players combined). Understanding these limitations is crucial in order to meaningfully interpret the results. 

%\michael{Consider adding something about unseen high-impact outliers being unlikely due to the correlation between publicity and value. That could be a good point for the stability of such queries. I think many reviewers might otherwise be sceptical of the ease of doing an average}

\vspace{-2pt}
\section{Experiments} \label{sec:eval}
We evaluated our  algorithms on several crowdsourced and synthetic data sets to test their predictive power. 
Crowdsourcing allowed us to generate many real data sets and avoided the licensing issues which often comes with other data sources.
We designed our experiments to answer the following questions:

\begin{itemize}
\item How does the estimation quality between the different estimators compare on real-world data sets? 
\item What is the sensitivity of our estimators in regard to data skew ({\em publicity-value correlation}) and streakers/imbalance of data sources?
\item How useful is the upper bound?
\item How early are accurate {\em MIN}/{\em MAX} estimates possible?
\end{itemize}

\vspace*{-4pt}
\subsection{Real Crowdsourced Data}\vspace*{-1pt}
We evaluated the estimation techniques on a number of real-world data sets, each gathered independently using Amazon Mechanical Turk, following the guidelines in \cite{crowddb}.
Here we chose four representative data sets and four aggregate queries, which show different characteristics we encountered during the evaluation.
\begin{enumerate}
\item {\bf US tech revenue \& employment}: For the query: {\em how much revenue does the US tech industry produce?}, i.e., \texttt{SELECT SUM(revenue) FROM us\_tech\_ companies}, we used the crowd to collect US\footnote{\scriptsize We asked for companies in Silicon Valley to get a representative sample of US tech companies; without restrictions we received too many tiny computer shops and even non-US based companies.} tech company names and revenues. Similarly, in an independent experiment we asked for US tech company names and number of employees, in order to answer the question: {\em how many people does the US tech industry employ?}, i.e., \texttt{SELECT SUM(employees) FROM us\_tech\_companies}. We selected the two data sets as they exhibit a steady arrival of unique answers from crowd workers. 
\item {\bf US GDP}: As a proof-of-concept experiment, we asked crowd workers to enter a US state with its GDP. This data set suffered from streakers.
\item {\bf Proton beam}: 
Together with researchers from the field of Evidence Based Medicine (EBM) (group-name omitted for double blind reviewing) we created a platform for abstract screening and fact extraction and spent over \$6,000 on AMT, to screen articles about 4 different topics. 
Here we utilize the results on one of these, namely Proton beam: a set of articles on the benefits and harms of charged-particle radiation therapy for patients with cancer. Part of the abstract screening asked workers to supply the number of patients being studied. The question we aim to answer is {\em how many people, in total, participated in these type of studies}: \texttt{SELECT SUM(participants) FROM proton\_beam\_studies}. 
This data set and research question is grounded in a real world problem and unlike the other queries, this one does not have a known answer. 
\end{enumerate}

\begin{figure}[t]
 \centering
 \includegraphics[width=2.7in]{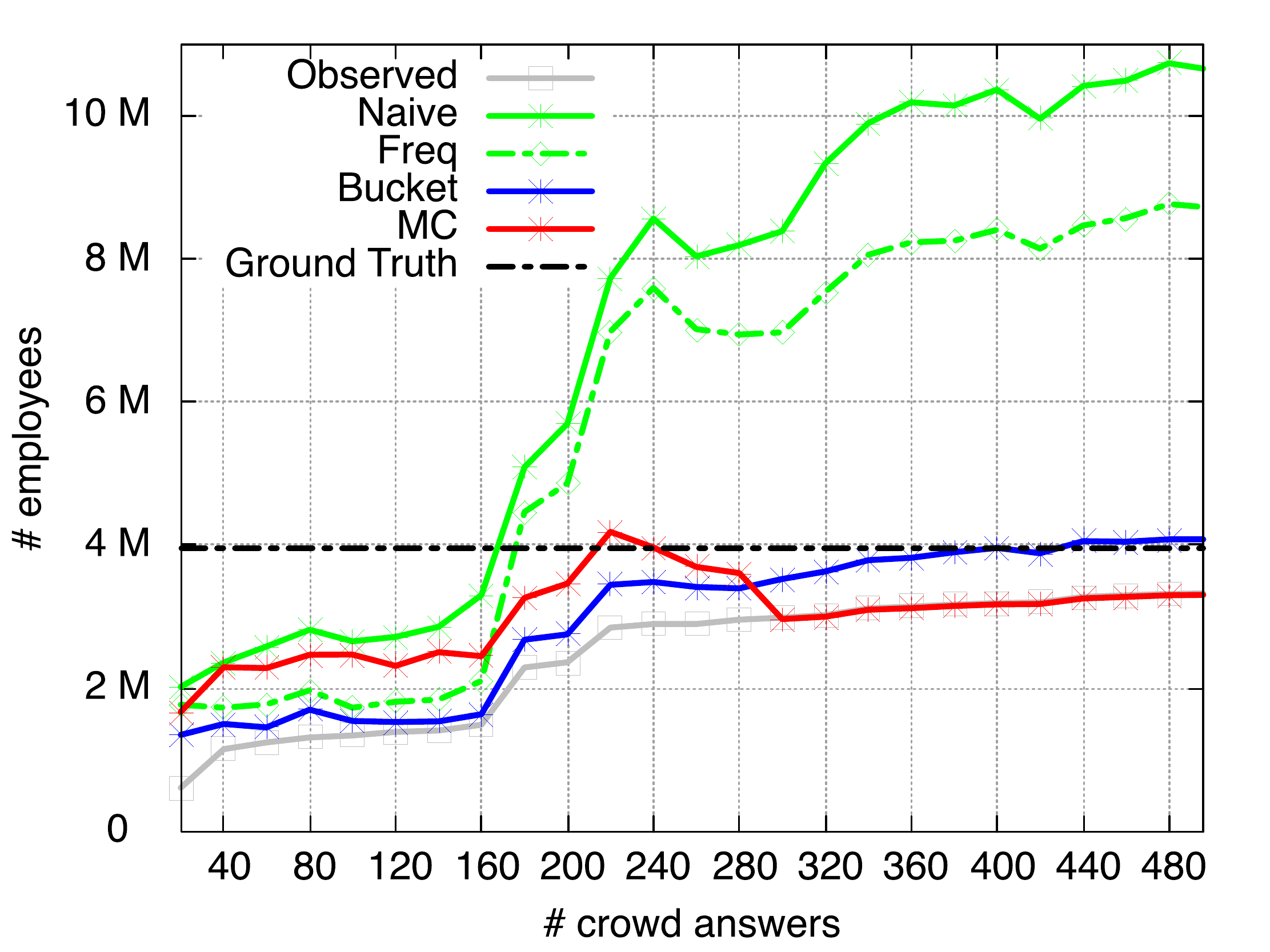}
 \caption{The best US tech-sector employment}
 %\vspace*{-18pt}
 \label{fig:real_employee}
\end{figure}

We paid between 2 and 35 cents per task.
For the Proton beam experiment we designed a qualification test and introduced hidden control tests to filter out bad workers ( reference is omitted for double blind reviewing), the other experiments were done without qualification tests. 
For the purposes of this study, we performed data cleaning manually:  
if workers disagreed on the value (e.g., the number of employees of a company) we used the average. 

In the following we describe the results for every data set and the following estimators: {\em \Naive{}} (naive) (Section~\ref{sec:sum:naive}), {\em frequency} (Freq) (Section~\ref{sec:sum:freq}),  {\em bucket} (Bucket) (Section~\ref{sec:sum:bucket}), and {\em Monte-Carlo} (MC) (Section~\ref{sec:sum:mc}) estimators (other estimators did not perform that well or had the same performance and are only shown in  Appendix~\ref{appendix:static} and \ref{appendix:other_estimators}).

\vspace*{-5pt}
\subsubsection{US Tech-Sector Employment}\label{sec:eval:employee}\vspace*{-1pt}
Figure~\ref{fig:real_employee} shows the {\em SUM} estimates from the different estimators (colored lines) for our running example \texttt{SELECT SUM(employees) FROM us\_tech\_companies} as well as the observed {\em SUM} (grey line) over time (i.e., with an increasing number of crowd-answers).
As the ground-truth (dotted black line) we used the US tech sector employment report from the Pew Research Center \cite{real_data_empl}.

Both the {\em \naive{}} and {\em frequency} estimators heavily overestimate the impact of {\em unknown unknowns}. 
The {\em frequency} estimator does slightly better than the {\em \naive{}} estimator, which indicates that some big companies have a high publicity likelihood and were observed early on by several sources.

\begin{figure*}[!t]
 \subfigure[US Tech Revenue]{\includegraphics[width = 2.3in]{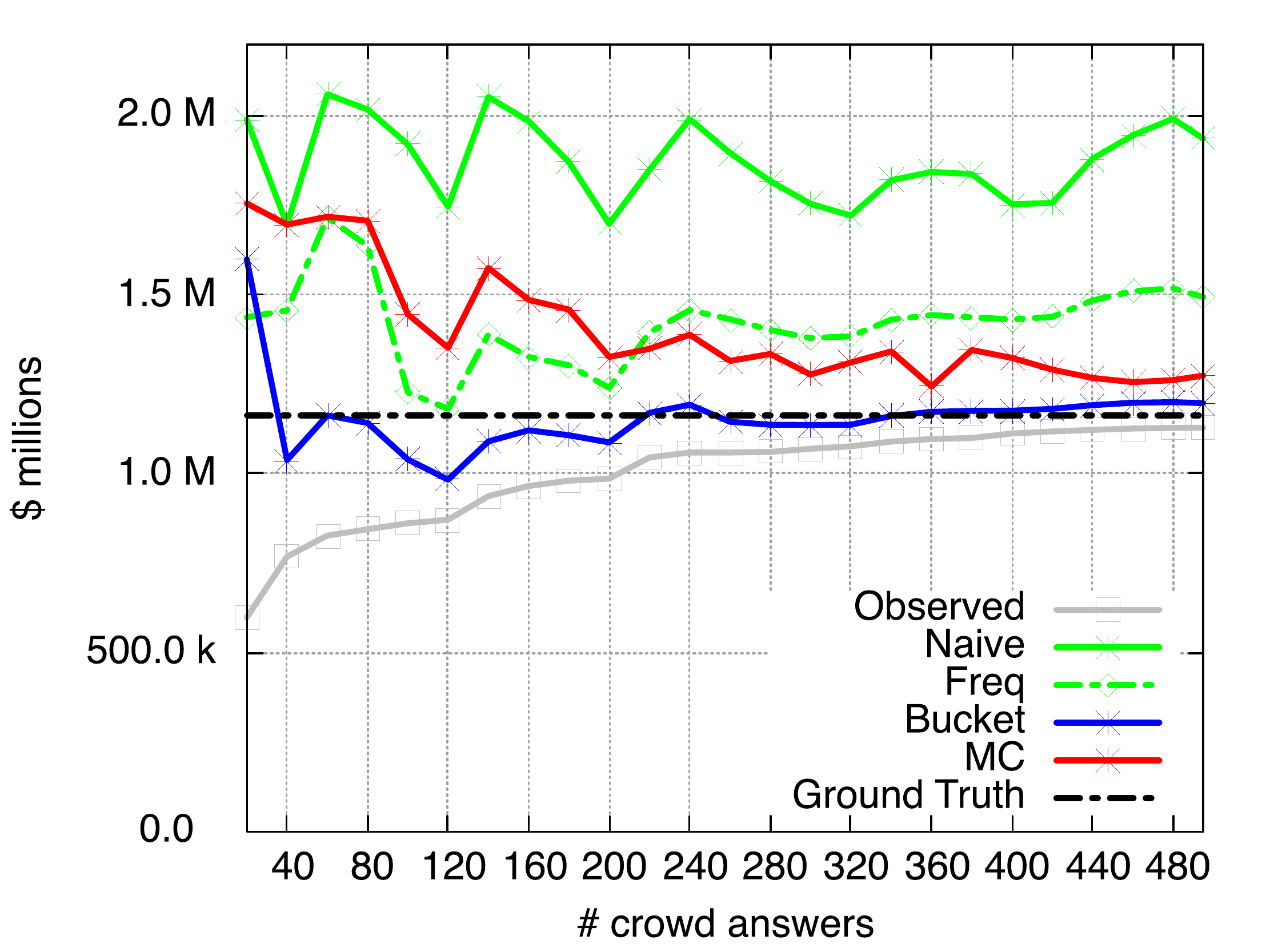}}%
 \subfigure[US GDP]{\includegraphics[width = 2.3in]{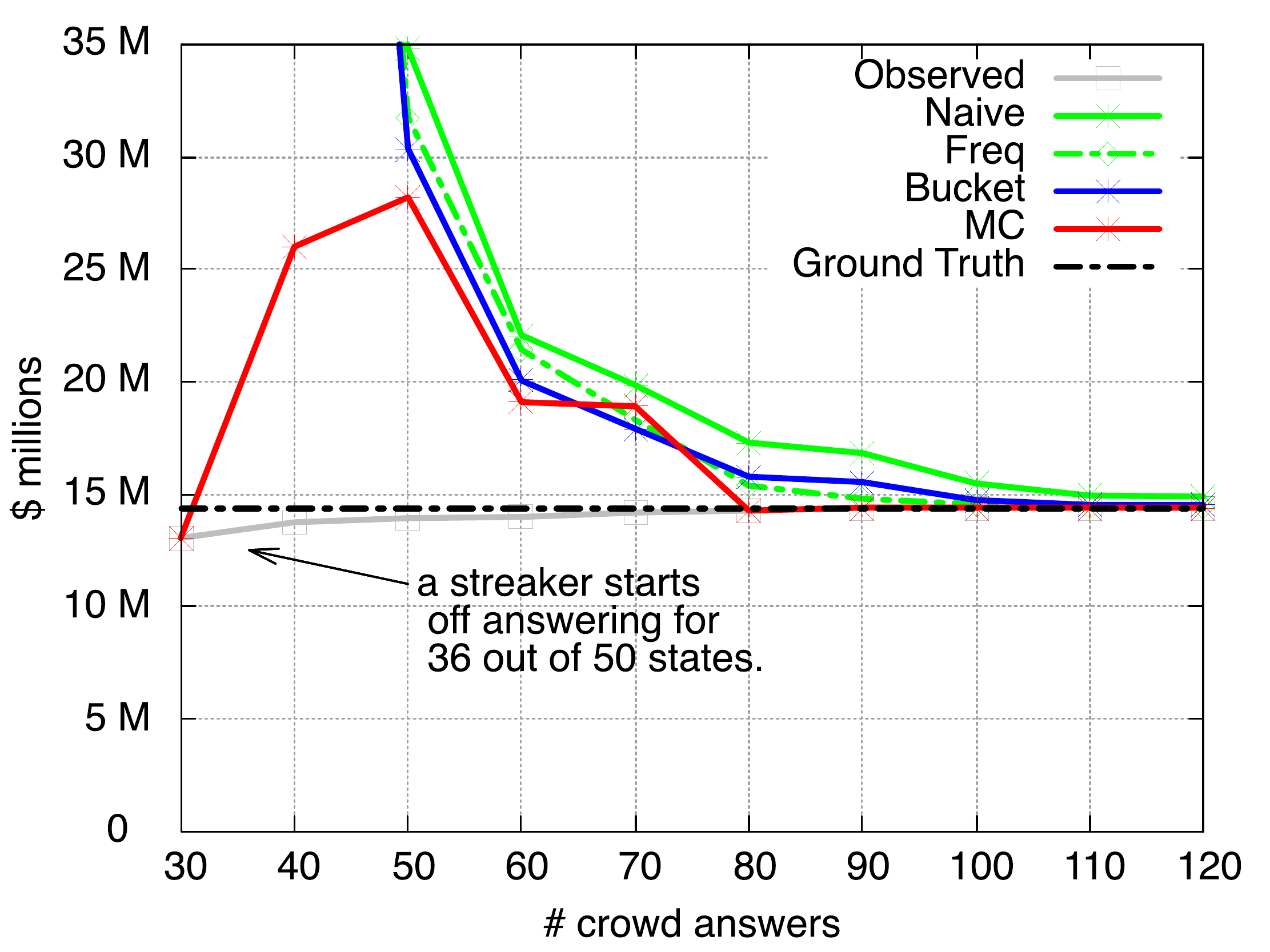}}%
 \subfigure[Proton Beam]{\includegraphics[width = 2.3in]{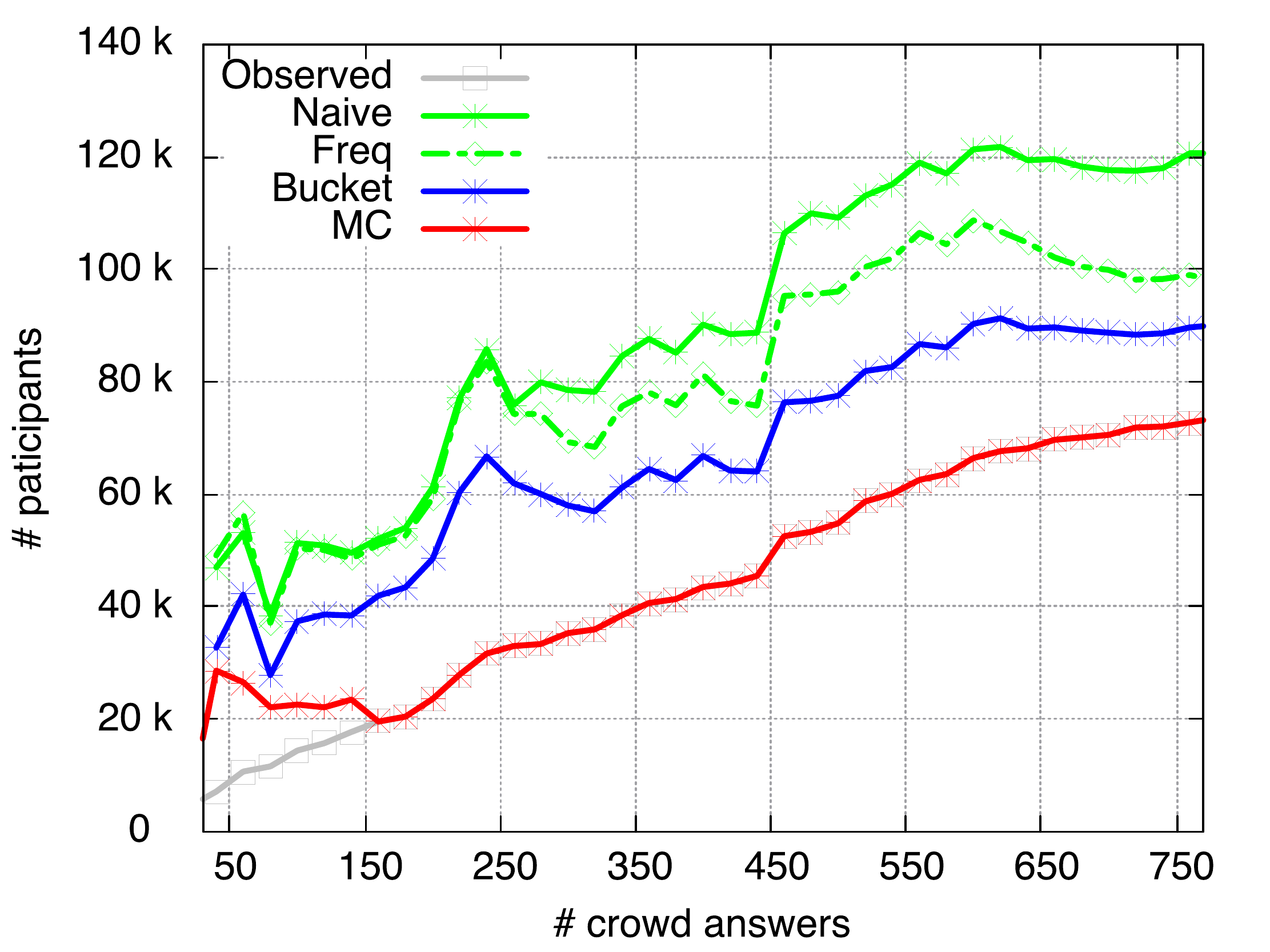}}
   \vspace*{-2pt}
 \caption{Real data experiments with aggregate SUM query}
  \vspace*{-16pt}
 \label{fig:real_source}
\end{figure*}

In contrast, the {\em MC} estimator does well until it falls back to the observed query result. 
This can be explained by a peculiarity of this experiment. 
After roughly 280 crowd-provided data item, all remaining companies have a rather uniform publicity likelihood.  
In such a case, the {\em MC} estimator has a tendency to favor count estimates, which are similar to the number of observed items: $\hat{N}_{MC} \sim c$. 
A major drawback of our {\em MC} estimation technique. 
%Note that when the shapes of distributions are as plain as uniform, any unmatched unique items in $Q$ ($\hat{N}_{MC} > c$) will penalize and increase the distance between $S$ and $Q$. 
%This is indeed a major drawback of our MC estimation technique and is an interesting area for future work. 

Finally, the {\em bucket} estimator provides the best estimate ($4053160.57$ at $500$ crowd answers),
which is only $\sim2.5\%$ above the ground truth ($3951730$). While it is possible that the {\em bucket} estimator might require more data to converge, it is also possible that the ground truth is inaccurate: the employment statistics can vary widely based on many factors (e.g., inclusion of part-time employees, tech sector definition). We also speculate that there exist many smaller US tech start-ups that might be overlooked by survey research agencies, due to the high data collection cost. In contrast, a school of crowd workers can more easily find smaller start-ups and their number of employees on web-pages. Thus, the  {\em bucket} estimate could be closer to the ground truth than the one by the Pew Research Center. This is an astonishing result as the cost of crowdsourcing (e.g., $\$50.00$ per 500 crowd-answers for US tech revenue \& employment experiments) is probably only a small fraction of the cost of survey research by any major agency.

\vspace*{-4pt}
\subsubsection{US Tech-Sector Revenue}\vspace*{-1pt}
Figure~\ref{fig:real_source}(a) shows the results for the US tech-sector revenue. In this data set, both the {\em \naive{}} and the {\em frequency} techniques overestimate the ground truth significantly because of the {\em publicity-value correlation}. 
While both estimators will eventually converge to the ground truth, it requires significantly more crowd-answers than what we collected. 
%As in the previous example, the {\em frequency} estimator performs better than the {\em \naive{}} estimator, which indicates that the really large US companies were observed early. 

Again, both {\em Monte-Carlo} and {\em bucket} estimators provide better estimates than {\em \naive{}} and {\em frequency} estimators .Yet, {\em Monte-Carlo} still overestimates, whereas {\em bucket} gives an almost perfect estimate after 240 answers.
However, it can also be observed that the bucket estimator slightly over-estimates at the end of the experiment. 
This happens because one crowd-worker suddenly reported a few unique smaller companies causing the estimator to believe that there were more. 
Again, we cannot say with 100\% certainty that our assumed ground-truth is actually the real ground truth and the {\em bucket} estimate might or might not be the real value.

\vspace*{-4pt}
\subsubsection{GDP per US State}\vspace*{-1pt}
\label{sec:eval:real:gdp}
Figure~\ref{fig:real_source}(b) shows the estimate quality for our GDP experiment.
To clean the data, we substituted the crowd reported GDP values with the values from \cite{real_data_gdp}. 
This experiment suffered from streakers, i.e., uneven contributions from crowd workers. 
A single crowd-worker reported almost all answers in the beginning; this kind of aggressive behavior results in unusually high $f_1$, which throws off the estimators. 

As the figure shows, only the {\em Monte-Carlo} based technique can actually deal with streakers and provides a reasonable estimate even in the beginning. 
However, it should also be noted that all estimators converge after 60 samples (for $N=50$).
Furthermore, except for the {\em Monte-Carlo} estimator, there is no difference between the other estimators.

\vspace*{-4pt}
\subsubsection{Proton Beam}\vspace*{-1pt}
Finally, results for Proton beam are shown in Figure~\ref{fig:real_source}(c).
Again the {\em Monte-Carlo} estimator follows the observed line, which makes the estimates less interesting. Furthermore, we suspect that the {\em \naive{}} and the {\em frequency} estimators overestimate with constantly increasing number of unique data items (reviewed articles). By manually examining the data set, we confirm that this crowdsourcing experiment did not encounter any streakers, which may cause our estimators (e.g., {\em bucket}) to fail. Note that the {\em bucket} estimator converges to roughly $95k$, which we consider to be the best estimate of the number of participants for this particular type of cancer therapy effectiveness study. 

\vspace*{-4pt}
\subsubsection{Discussion}\vspace*{-1pt}
Overall, our {\em bucket} estimator has the highest accuracy.
The only exception is when streakers are present, making the {\em Monte Carlo} to perform better. 
However, it should also be noted, that the run-time of the {\em Monte-Carlo} estimator is significantly higher than the other estimators.
While not a serious issue for our experiments (roughly 3.5s for {\em Monte-Carlo} vs. 0.2s for {\em bucket}), it could be significant for larger data sets, as the run-time scales linearly with sample size (the inner loop in Algorithm~\ref{alg:mc_method} depends on the sample size).
In the remainder we analyze the different estimators in more depths using simulation and make final recommendations about which estimator to use at the end of the section.

\begin{figure*}[!t]
 \centering
 \includegraphics[width=\linewidth,height=10cm]{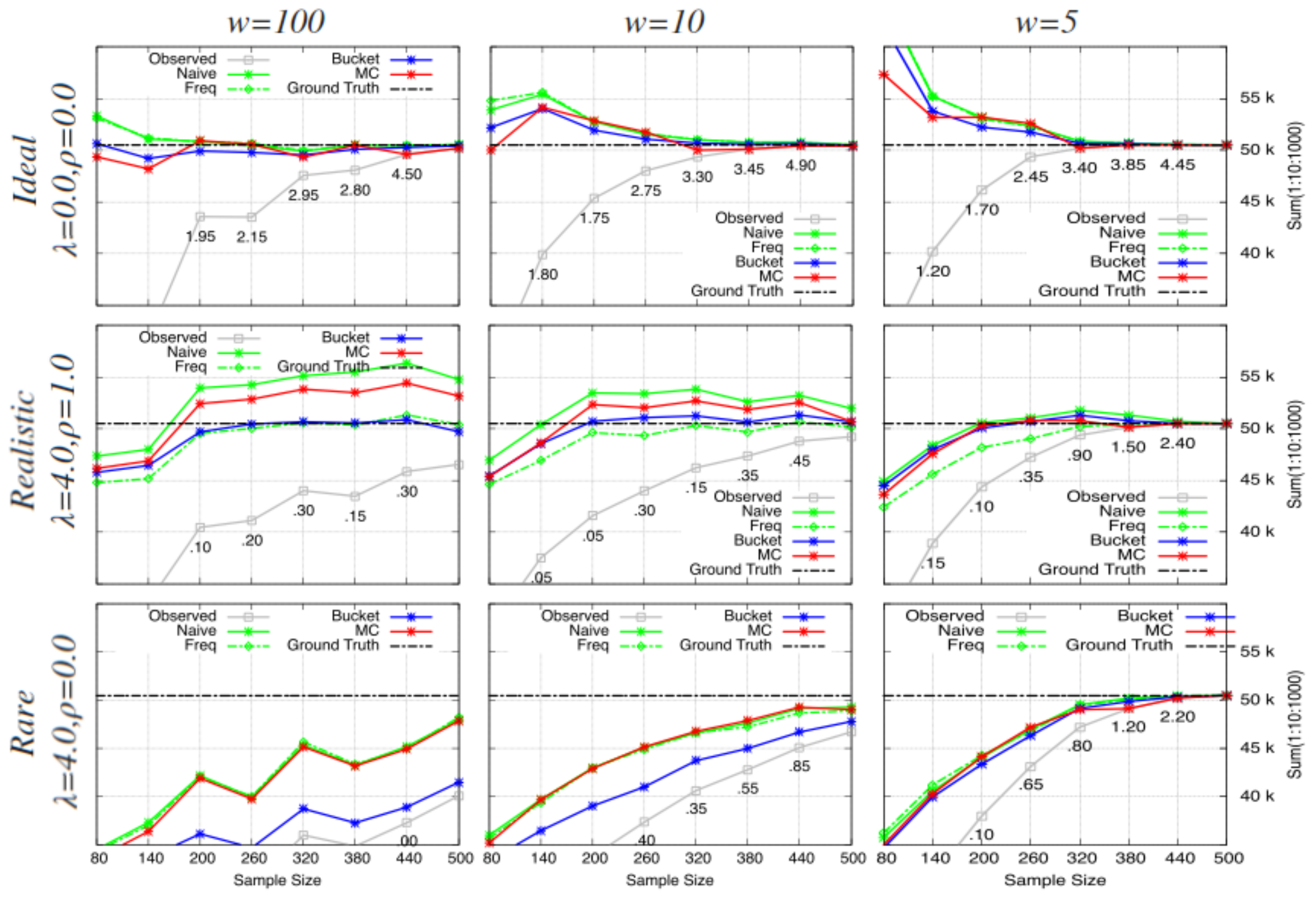}
 \caption{Synthetic data with varying number of sources ($w$), degrees of publicity skew ($\lambda$) \& publicity-value correlation ($\rho$).}
 %\vspace*{-18pt}
 \label{fig:synt_sum}
 \vspace*{-13pt}
\end{figure*}

\subsection{Synthetic Data Experiment}
\label{sec:exp:synt}

To explore the estimation quality more systematically, we used a  synthetic data set with $N=100$ unique items, each having a single attribute-value ranging from $10$ to $1000$ ($attr={10,20,30,...,1000}$). We further simulated the sampling process outlined in Section~\ref{sec:problem} and used an exponential distribution with parameter $\lambda$ to model various publicity distributions ( $\lambda=0$: uniform; $\lambda=4$: highly skewed). 
Finally, our simulation allowed us to vary the {\em publicity-value correlation} ( $\rho=0$: no correlation; $\rho=1$: perfect correlation - the most frequent item also has the largest value).

Figure~\ref{fig:synt_sum} shows the results for various synthetic data experiments, each of which is repeated 50 times and the results averaged (we omit the error bars for better readability). 
From left to right, we vary the number of simulated crowd-workers (i.e., sources) from $w=100$, $10$ to $5$. 
From top to bottom, we first assume no publicity skew and no {\em publicity-value correlation} ($\lambda=0, \rho=0$), a for species estimation techniques often ideal scenario, we then show the more realistic scenario with skew and publicity-value correlation ($\lambda=4, \rho=1$), and finally simulate an environment where some rare items might contain high values ($\lambda=4, \rho=0$).

{\bf Ideal:}
Looking at the top-left figure with a uniform publicity distribution and a hundred workers, we can see that all estimators perform very well from the beginning. 
This is not surprising as all estimators work best with sampling with replacement from a uniform publicity distribution; having many workers sampling without replacement from a uniform distribution approximates sampling with replacement. 
With fewer numbers of workers sampling from the uniform distribution (top row), all estimators start to overestimate slightly.
We conclude, that {\em under the ideal conditions (i.e., the original assumptions of species estimation technique)  all estimators perform equally well.}

{\bf Realistic:}
The middle row shows the scenarios which best resemble real-world use cases as it considers a skewed {\em publicity} distribution with  a positive {\em publicity-value correlation}.
In this case, the {\em bucket} estimator always provides the best estimates.
However, in contrast to the real-world experiments the {\em frequency} estimator also performs well. 
This is due to a couple of reasons: Firstly, the {\em publicity} is highly skewed and perfectly correlated to the values. Secondly, the item values are evenly spaced.
This helps the {\em frequency} estimator to under-estimate as {\em singletons} consist of only rare low-valued items from the tail -- a peculiarity of this simulation. 
Also interestingly, with 5 evenly contributing workers almost all estimators perform about the same. 
However, the {\em bucket} estimator has less variance (not shown). 
We conclude, that {\em under the more realistic conditions the bucket estimator performs the best and does not over-estimate the value}.
%This can be explained by the high publicity skew and the fact that sources are sampling without replacement; consequently, all unique values appear quickly more than once in $S$, creat

%this allow for less room for improvement (i.e., almost no singletons) and thus, cause similar estimates.  

{\bf Rare events:}
Finally, we see in the bottom row that the {\em bucket} estimator is not the best choice. This is the case where we have skewed {\em publicity}, but no {\em publicity-value correlation}. 
In fact, all estimators perform poorly in this scenario, even with a lot of data sources (d).
As the {\em publicity} distribution tail can take on any values (i.e., no {\em publicity-value correlation}, the tail (i.e., {\em singletons}) can contain many high-impact values or ``black-swan'' events. In this case, because it conservatively favors underestimation, the {\em bucket} estimator performs worse.
In summary, {\em none of the estimators are able to predict black-swan events or the long tail; all the estimators underestimate the ground truth}.

\begin{comment}
\begin{figure*}[!t]
 \vspace*{-1em}\subfigure[$w=100$,$\lambda=0.0$,$\rho=0.0$]{\includegraphics[width = 2.3in, height=1.1in]{figures/w100_uniform_unc.pdf}}%
\subfigure[$w=10$,$\lambda=0.0$,$\rho=0.0$]{\includegraphics[width = 2.3in, height=1.1in]{figures/w10_uniform_unc.pdf}}%
 \subfigure[$w=5$,$\lambda=0.0$,$\rho=0.0$]{\includegraphics[width = 2.3in, height=1.1in]{figures/w5_uniform_unc.pdf}}%
 \\
 \subfigure[$w=100$,$\lambda=4.0$,$\rho=0.0$]{\includegraphics[width = 2.3in, height=1.1in]{figures/w100_skewed_unc.pdf}}%
 \subfigure[$w=10$,$\lambda=4.0$,$\rho=0.0$]{\includegraphics[width = 2.3in, height=1.1in]{figures/w10_skewed_unc.pdf}}%
 \subfigure[$w=5$,$\lambda=4.0$,$\rho=0.0$]{\includegraphics[width = 2.3in, height=1.1in]{figures/w5_skewed_unc.pdf}}%
 \\
 \subfigure[$w=100$,$\lambda=4.0$,$\rho=1.0$]{\includegraphics[width = 2.3in, height=1.1in]{figures/w100_skewed_cor.pdf}}%
 \subfigure[$w=10$,$\lambda=4.0$,$\rho=1.0$]{\includegraphics[width = 2.3in, height=1.1in]{figures/w10_skewed_cor.pdf}}%
 \subfigure[$w=5$,$\lambda=4.0$,$\rho=1.0$]{\includegraphics[width = 2.3in, height=1.1in]{figures/w5_skewed_cor.pdf}}%
 \caption{Synthetic data with varying number of sources ($w$), degrees of publicity skew ($\lambda$) \& publicity-value correlation ($\rho$).}
 \label{fig:synt_sum}
   \vspace*{-10pt}
\end{figure*}
\end{comment}

\subsection{Streakers}\label{sec:streaker}
We have seen in Section~\ref{sec:eval:real:gdp} that the estimators can heavily overestimate in the presence of streakers. We now examine the effects of streakers using the synthetic data set with $n=20$, $\lambda=1.0$ and $\rho=1.0$.

\begin{figure*}[!t]
 \subfigure[streakers only]{\includegraphics[width = 2.3in, height=1.1in]{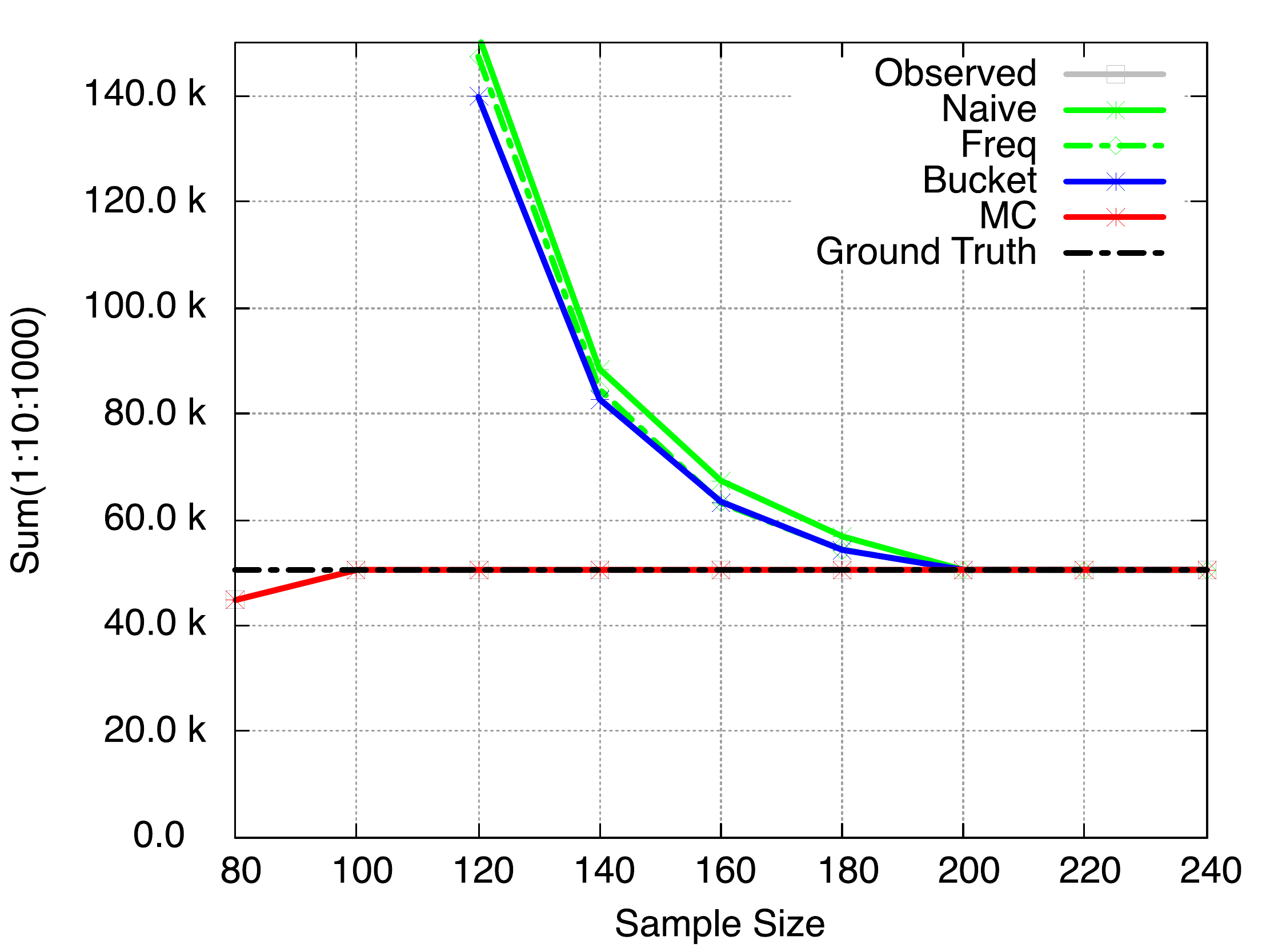}}
 \subfigure[a streaker injected at $n=160$]{\includegraphics[width = 2.3in, height=1.1in]{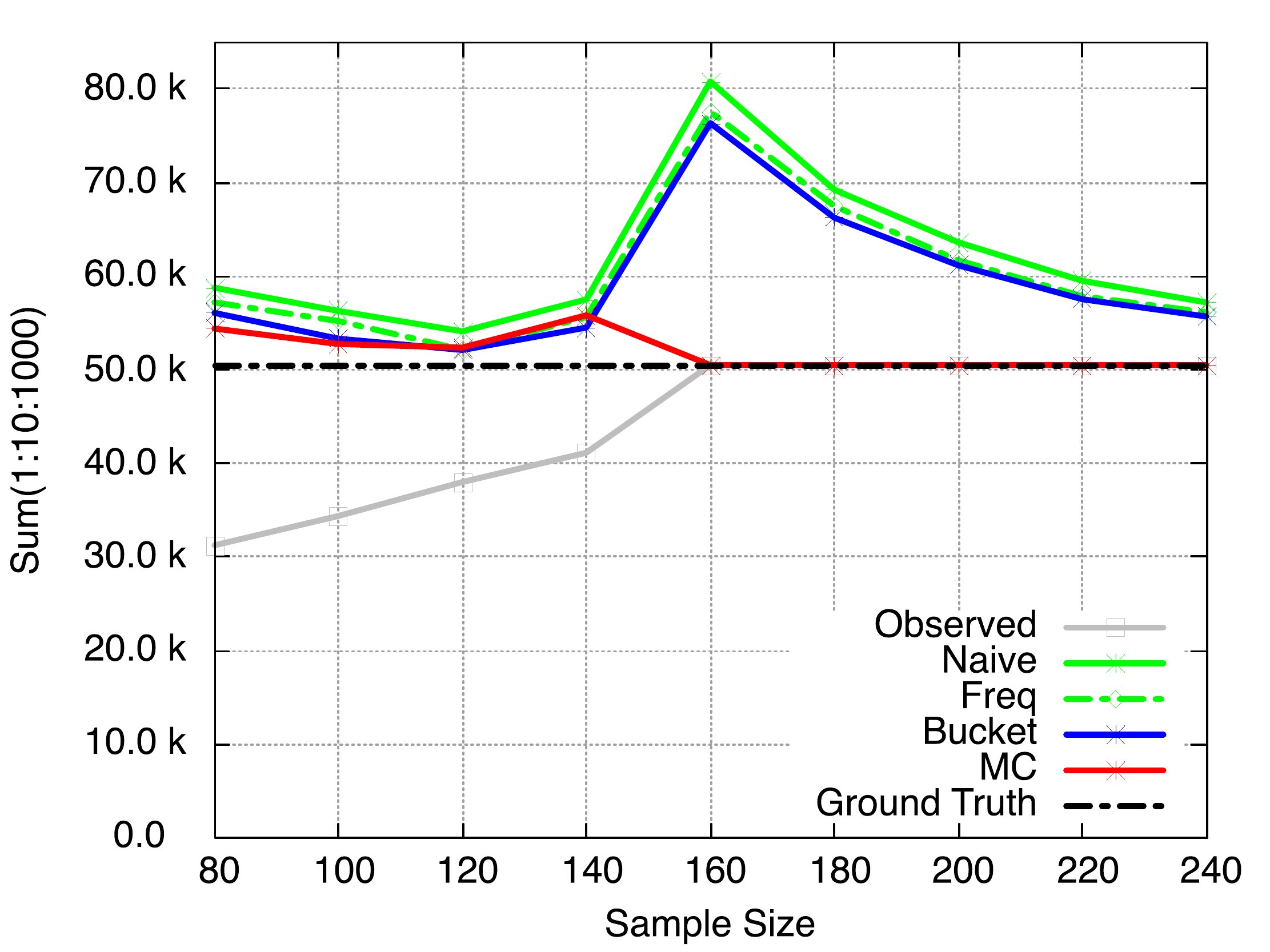}}
 \subfigure[upper bound]{\includegraphics[width = 2.3in, height=1.1in]{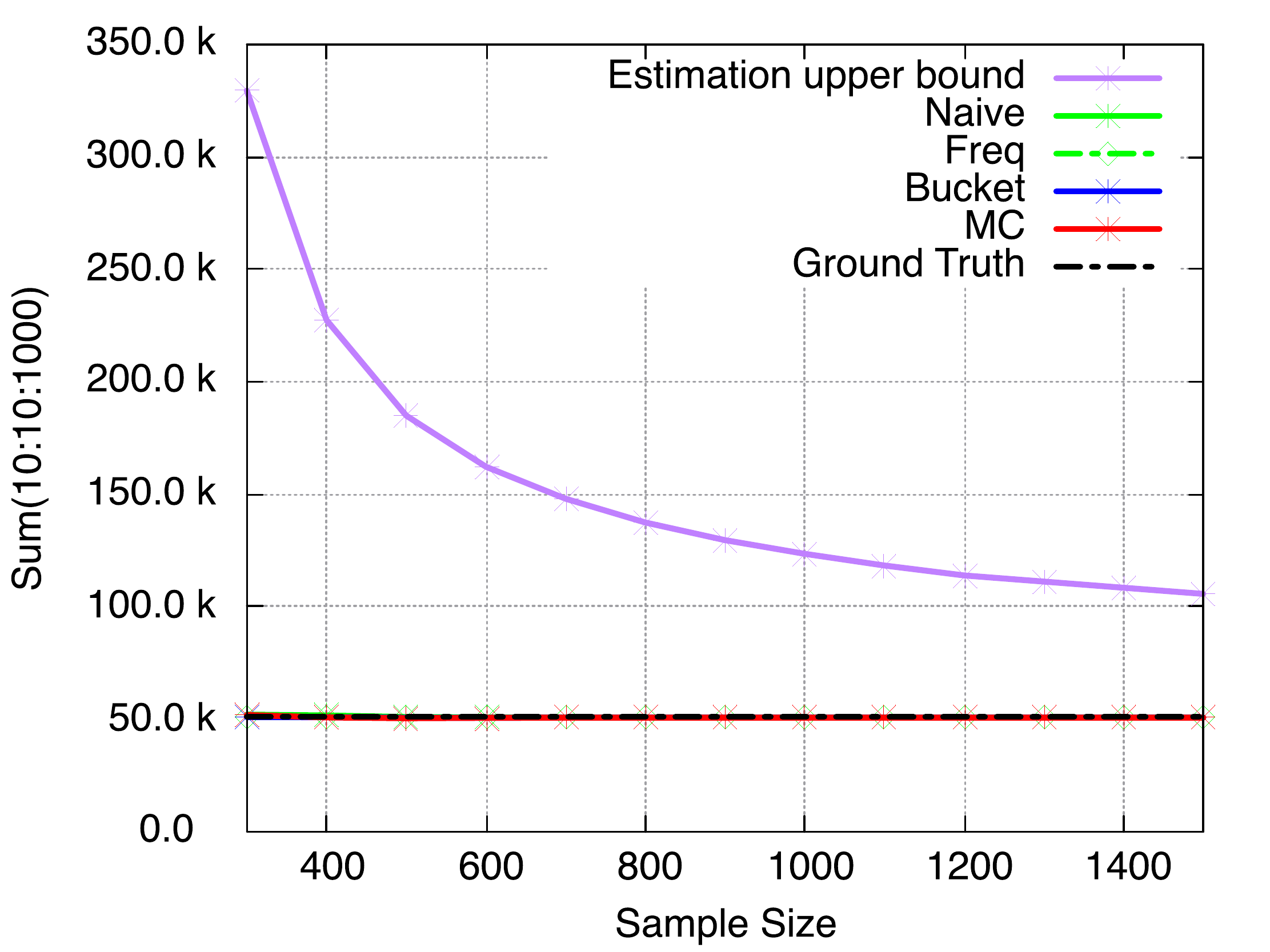}} \vspace*{-10pt}
 \\ 
  \subfigure[AVG]{\includegraphics[width = 2.3in, height=1.1in]{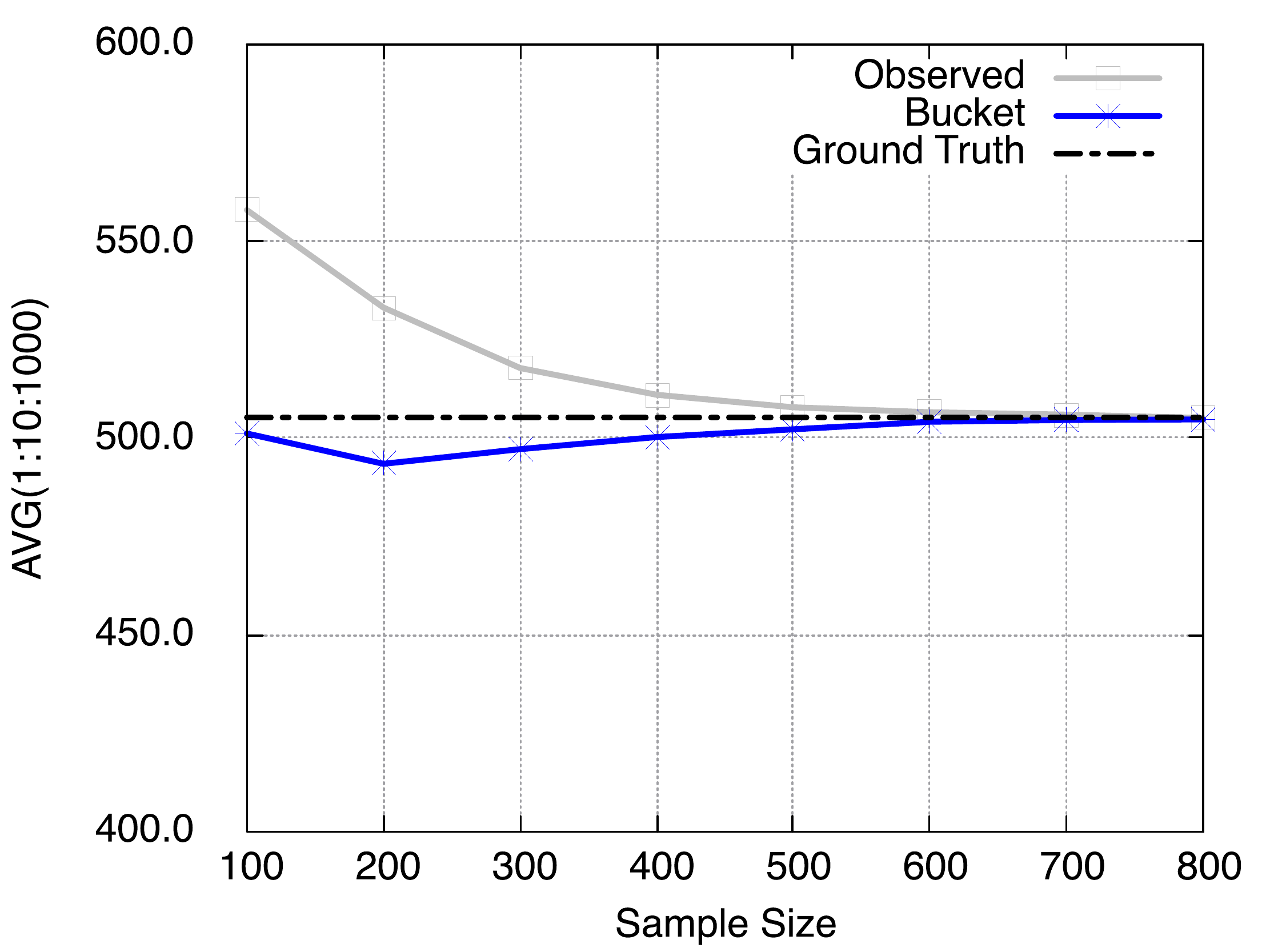}}%
 \subfigure[MAX]{\includegraphics[width = 2.3in, height=1.1in]{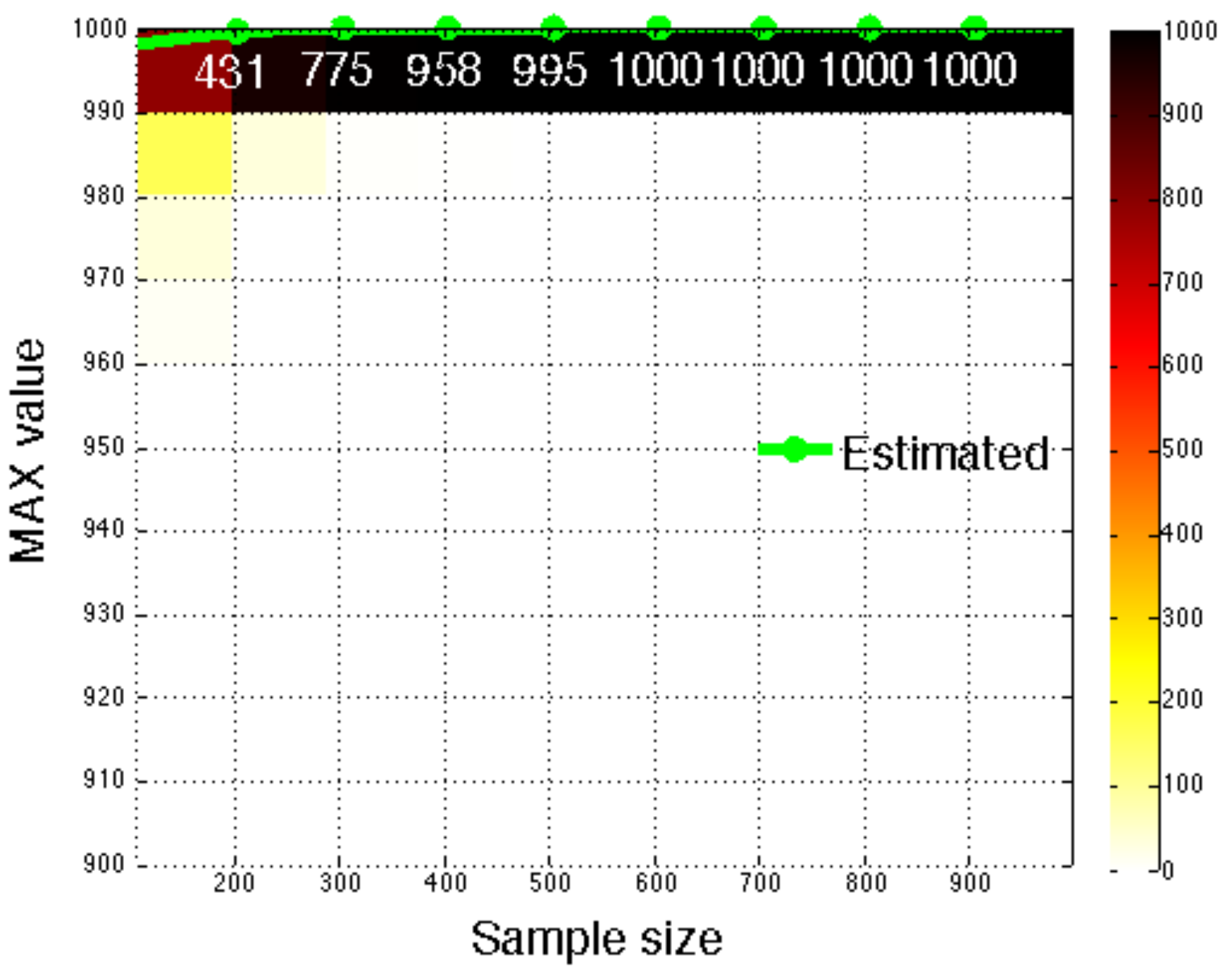}}\hspace{0.3em}%
 \subfigure[MIN]{\includegraphics[width = 2.3in, height=1.1in]{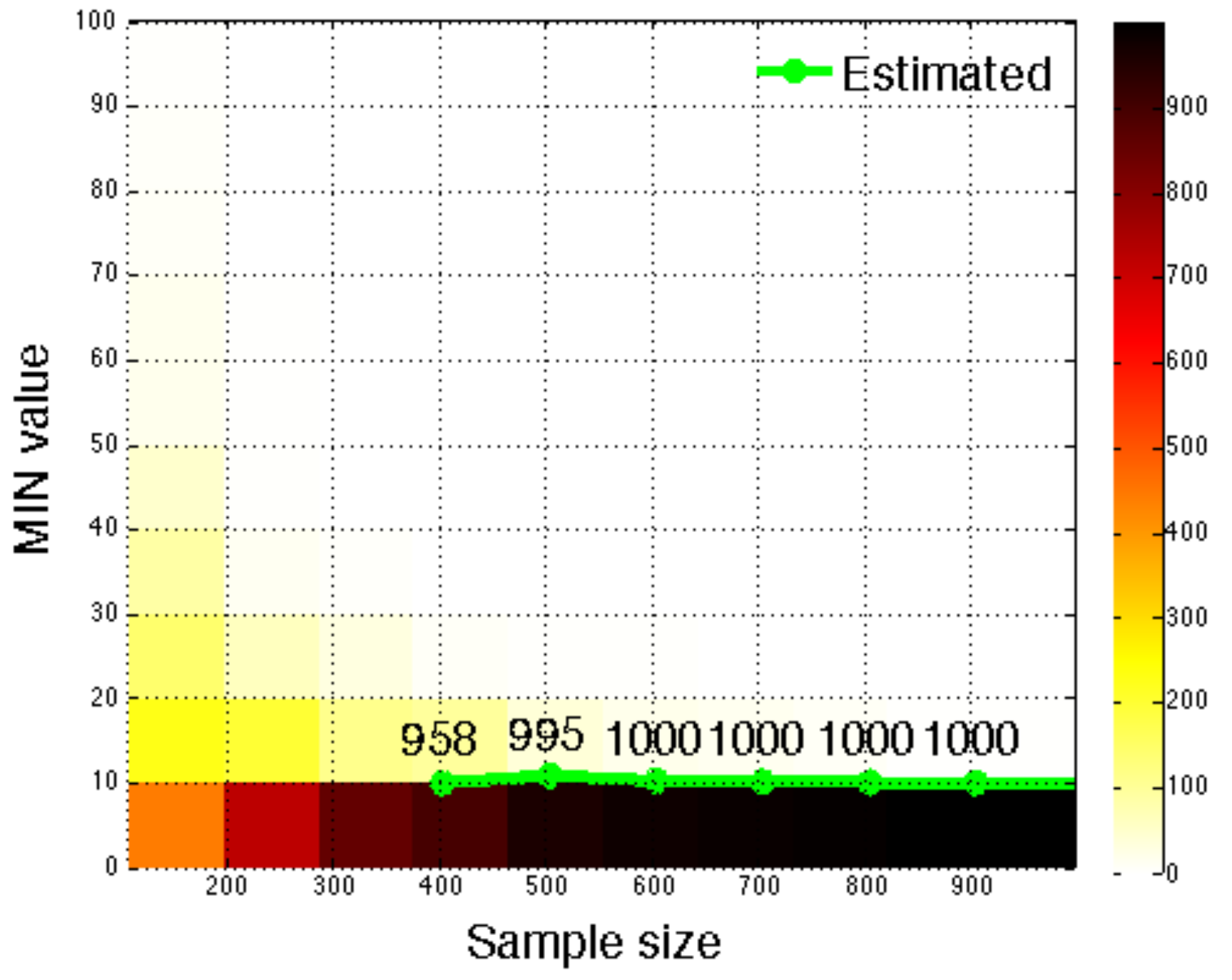}}\hspace{0.3em} \vspace*{-5pt}
 \caption{Streaker effect (a-b), estimation upper bound (c), AVG query (d) and aggregate MAX/MIN queries (e)(f) experiments using a synthetic data ($\lambda=1.0$, $\rho=1.0$: larger values are more likely)}
 \label{fig:synt_streaker_other}
 \vspace*{-12pt}
\end{figure*}

First, we consider an extreme case where each source successively provides all $N=100$ data items; first, one data source contributes $n=100$ items and then the second source starts to contribute its $n=100$ items, and so on. Figure~\ref{fig:synt_streaker_other}(a) shows that {\em Monte-Carlo} simply defaults to the observed sum from one source ($n=100$), whereas all other estimators fail. This is because of the fact that all $Chao92$-based estimators assume a sample with replacement; an assumption which is strongly violated in this case. 
Only {\em Monte-Carlo} is more robust against streakers as it tries to best explain the observed $S$ using simulation. 
%Also note that the Bucket estimators perform similar to the Naive and Freq estimators. This is because the {\em publicity} is not skewed heavily and the sample $S$ appears to be more uniform.

Next, we consider a more moderate case where we inject a single streaker (i.e., an overly ambitious crowd-worker). In Figure~\ref{fig:synt_streaker_other}(b) a streaker is injected at the sample size $n=160$, contributing all $N=100$ unique data items directly afterwards. 
Similar to the previous case, all estimators, except {\em Monte-Carlo}, heavily overestimate in the presence of a streaker.
Again, the reason is that {\em Monte-Carlo} uses simulation to explain the observed sample $S$ instead of assuming that $S$ was created using sampling {\em with} replacement.

\subsection{Other Queries \& Upper Bound}\label{sec:eval_other}
In this subsection we present results for other aggregate queries than {\em SUM} using the techniques from Section~\ref{sec:other_query}. 
As before we use synthetic data with 100 unique data items (e.g., with values $\{10,20,30,...,1000\}$) integrated over 20 sources with $\lambda=1.0$ and a publicity-value correlation $\rho=1.0$. The experiments are repeated 1000 times.

{\bf AVG:} Figure~~\ref{fig:synt_streaker_other}(c) shows the observed (gray line) and estimated (blue line) for a simple average query of the form \texttt{SELECT AVG(attr) FROM table}. We only show the {\em bucket} estimation, as other estimates exactly overlap the observed \emph{AVG} query results (i.e., when all {\em unknown unknowns} assume the same observed mean value, the {\em AVG} query result is the same as the observed). As with the sum-aggregates, our dynamic bucket estimator is able to correct the bias of the average because of the publicity-value correlation and provides an almost perfect estimate in this scenario.

{\bf MIN/MAX:} Figure~\ref{fig:synt_streaker_other}(d-e) compactly visualizes the observed \em{MIN} or \em{MAX} query results.
The heat-map shows when the real MIN/MAX value was observed in the data set (the darker the color the more often the result was observed given a number of samples over the 1000 repetitions). 
The green line shows on average, which value was reported if the  {\em unknown unknowns} count estimate for the highest ({\em MAX}) / lowest ({\em MIN}) bucket was zero. 
The text next to the green line shows how often over the 1000 repetitions the MIN/MAX value was reported for a given sample size.
As it can be seen the average is almost perfect for both {\em MAX} and {\em MIN} (note the actual minimum value is 10). 
That is, whenever our estimation technique for {\em MAX}/{\em MIN} reports a value the user can have more trust in it. 
It should be noted tough, that it is impossible to estimate rare extreme values (black swans). 
Thus, it is only possible to improve upon the confidence but not eliminate any doubts in the results.

{\bf Upper Bound:}
Finally in Figure~\ref{fig:synt_streaker_other}(f) we show the upper-bound from Section~\ref{sec:upper_bound} using the same synthetic data set.
As it can be seen, the bound is very loose (i.e., very large compared to our estimates) and becomes more tight as we observe more data. 
We observed the same behavior over the real-world data sets (omitted due to space constraints).
While the upper bound provides a valuable insight, it may still be too loose for many real-world scenarios and we hope to improve it in the future. 

\vspace{-8pt}
\subsection{Summary}\vspace{-1pt}
\label{sec:eval:summary}
{\bf Which Estimator To Use}
While the {\em Monte Carlo} or {\em bucket} estimators always dominate all the others, there is no clear winner between them. 
The {\em bucket} estimator performs exceptionally well  unless the data sources are imbalanced. 
It provides the best performance on the real-world use cases (except on the GDP experiment, which suffers from streakers); furthermore, it performs at least as good as other estimators in the simulations from Section~\ref{sec:exp:synt} (except for the {\em rare event} case, in which all estimators fail to predict black-swan events).
However, when the data sources are imbalanced the {\em Monte Carlo} estimator wins. 

%While this suggests that it might be possible to automatically switch between the  approaches, we neglected the idea as we believe it is important that the analysts understands the different meaning of the underlying  statistical approaches. 
%Though it should be noted, that we tried to use the MC method within the bucket estimator instead of $Chao92$, which did not work well (see Appendix~\ref{appendix:other_estimators}).

%\yeounoh{introducing new (hard) terms without explanations; the paragraph below may arouse more questions + some disputes if we are not precise in describing them.  ALso, Chao92 -- sample coverage based method -- does not work well with skewed distributions, even with skewness correction; it works well with uniform dist, and our bucket approach guarantees somewhat more uniform dist per bucket.} \tim{Yes, but I also want to avoid this very soft descriptions about overlap. In addition, I want to make the point that the user needs not understand the statistical difference. I tried to combine both approaches}

The reason is, that the {\em bucket} estimator is a sample coverage-based method as it uses $Chao92$ and thus, a {\em nonparametric model}, which does not require assumptions about the underlying distribution.  
However, it assumes a single sample without replacement. 
This assumption is not an issue as long as enough independent data sources exists (using simulations we found that 5 sources are often sufficient, see Appendix~\ref{appendix:num_sources})  and every data source contributes evenly to $S$ (i.e., there are no streakers).
%Furthermore, while theoretically the $Chao92$ should be correct early on, in  \cite{chao92} the authors found that a sample coverage of 50\% is required to get good estimates.

In contrast, the {\em Monte-Carlo} estimator is a form of a {\em Data-Analytic Methods} and really good at adjusting to the specifically observed sampling scenario (i.e., streakers), but at a cost of being a {\em parametric model}.
The method assumes an exponential distribution to model the {\em publicity} distribution, which can be good or bad depending on the true shape of the underlying  distribution.
Thus, our recommendation is to use the {\em bucket} estimator, when the analyst knows that enough data sources contribute evenly to the sample, and, otherwise, to use the more conservative {\em Monte Carlo} method.

While theoretically the {\em bucket} estimator should be fairly accurate early on,  the authors of \cite{chao92} found that the $Chao92$ estimator is inaccurate with very low sample coverage $C$ (i.e., observed items are mostly {\em singletons}) and reported results for cases with $C \geq 0.395$ only. Based on that result, we make the general recommendation to use the estimates if the predicted sample coverage $\hat{C}$ (Equation~\ref{eqn:coverage}) is greater than $40\%$.

%{\em Monte-Carlo} estimator requires larger samples to more accurately simulate the population; however, splitting into buckets (i.e., using {\em bucket} estimator) generally means smaller sample per bucket. As a consequence, {\em Monte-Carlo} estimator is less accurate than other estimators per bucket and, therefore, overall. Furthermore, {\em Monte-Carlo} estimator assumes an exponential distribution to model {\em publicity}, which can be good or bad depending on the true shape of the underlying {\em publicity} distribution. This is not the case with {\em bucket} estimator, which uses {\em nonparametric} $Chao92$ and works well with any {\em publicity} distribution (e.g., uniform and skewed {\em publicity}) as shown in Section~\ref{sec:exp:synt}. 

%On the other hand, {\em bucket} estimator which implicitly assumes a sample sampled with replacement. This means that {\em bucket} estimator needs more number of sources contributing evenly to the sample (see Appendix~\ref{appendix:num_sources}) to have sufficient overlaps in the sample. Thus, our recommendation is to use {\em bucket} estimator, when enough data sources contribute evenly to the sample (i.e., there are no streakers), and, otherwise, to use the more conservative {\em Monte Carlo} method.

%Finally,  the {\em frequency} is  a great and simple technique to quickly test if a result is impacted by missing data though it never provides the best estimates. 

{\bf Trust In The Results}
With any types of estimators the main question arises: {\em How can we trust the estimate?}
In 1953, Good, who worked with Turing on the estimators, already pointed out that ``I don’t believe it is usually possible to estimate the number of species ... but only an appropriate lower bound to that number. This is because there is nearly always a good chance that there are a very large number of extremely rare species''\cite{bunge_review93}. 
%Harris puts it even more blunt and said that, "there is no way for the experimenter to establish the non-existence of an arbitrarily large number of classes each with negligible probability."\tim{please find the ORIGINAL citation}.
In estimating the Impact of {\em unknown unknowns}, this statement is even more critical as the rare items can have extreme values.

Yet besides this obvious risks and assumptions, species estimation techniques are extensively used in biology and even helped to decipher the Enigma machine \cite{turing53}. 
We actually believe that it comes down to a simple question: {\em What do you trust more? A potentially wrong answer as no missing data is considered or a potentially wrongly corrected result}.
Now knowing, that with enough sources and no imbalance of sources, our {\em bucket} estimator rather under- than over-estimates, it can generally be said that it can only improve the answer (see the simulations and real-world experiments). 
With imbalance of and/or only a few data sources, the answer is less clear, as the estimators  also more often over-estimate, even the conservative {\em Monte Carlo} technique (e.g., see Figure \ref{fig:real_source}(b)).
Thus, the true answer lies probably somewhere in between. 
With the help of our upper bound,  we can give the user at least a value range and an idea where the true value might be. 
It should be noted though, that the upper bound requires also two new assumptions: an item probability of at least  $1-\epsilon$ and that the value mean follows a normal distribution, which in some  rare cases might be violated. 
Still we believe, knowing something is wrong and a best guess, where the true value might be, is better than staying on the blind-side.
In this work, we made a first step in the direction, while a lot remains to be done from developing more tighter bounds, better ways to deal with the imbalance of sources, and easier ways to convey the meaning (and assumptions) of the estimates to the user.

%as heat maps (frequencies in color scale); the green lines represent the average estimates (i.e., report MAX/MIN values if and only if the {\em unknown unknowns} count estimation is zero in the highest/lowest value range bucket). The text labels on the green lines represent the number of times our estimator reported  \emph{MIN} or \emph{MAX} values. We see more reported extreme values as the sample size increases. As {\em publicity} is skewed to favor data items with larger values (e.g., a data item with the true minimum value $10$ is less likely to be sampled), we see that the observed MIN values are more dispersed. The estimation is more careful and report minimum values that are more likely to be correct. It is nearly impossible to estimate the extreme values exactly in the presence of {\em unknown unknowns}; however, our technique can still provide a hint to the correctness of the observed extreme values.

%- theoritcally it should work with any sample coverage but found that bigger than 30%, bigger than 50% chao92
% 

\section{Related Work} \label{sec:related}
Traditional  query processing assumes the database to be complete (i.e., closed world assumption). Furthermore, nearly all sampling-based query processing techniques assume knowledge of the population size \cite{haas}; hence, none of these are suitable for our problem with {\em unknown unknowns}. To the best of our knowledge, this is the first work on estimating the impact of the {\em unknown unknowns} on query results (i.e., aggregate query processing in the {\em open world}).

{\bf Species estimation:}
Most related to this work are the various species estimation techniques, like $Chao92$ \cite{chao92,chao84,bunge_review93}.
Recent work  \cite{valiant2011estimating} in this area even tries to estimate the  shape of the population (e.g., support size, $N$). 
We could use these techniques in place of $Chao92$ to estimate the number of {\em unknown unknowns}, but not to directly estimate the {\em impact of unknown}, as the shape does not concern the values of {\em unknown unknowns}.

Species estimation techniques have also been used to estimate the size of search engine indexes and the deep web \cite{DeepWebSurvey}. The problem is similar to our {\em unknown unknowns} count estimation, and the most common technique (i.e., capture-recapture) is also based on the species estimation techniques \cite{cc}. However, they again do not consider the {\em unknown unknowns} value. 

Species estimation techniques have also been used in the context of distinct  value estimation for a database table \cite{haas,charikar}.
However those techniques leverage the knowledge of the table size to avoid over-estimation.

{\bf Survey Methodology \& Missing Data:} \label{sec:survey}
There is a vast body of literature on sampling-based statistical inferences to estimate population statistics \cite{sapsford1999survey,mcclave2013statistics,kish1965survey} or techniques to deal with  missingness of values \cite{rubin1976inference,allison2012handling,dempster1977maximum,d2000estimating,yuan2010multiple}.

However, {\em unknown unknowns} are different from the missing data; missingness refers to the case when the record is known, but one (or more) of the values/attributes is missing. 
In addition, most of the techniques assume to know the population size to categorize something as missing (e.g., a registered subject participates and leaves before the study completes, a subject deliberately returns an empty questionnaire, only this many subjects out of that many people responded, etc.) and, to some extent, knowing the cause of missingness (e.g., missing completely at random, missing at random, missing not at random) to select appropriate techniques. Moreover, the statistical inference techniques, e.g., multiple imputation based EM/maximum likelihood estimation \cite{allison2012handling,dempster1977maximum}, propensity score estimation \cite{d2000estimating}, or {\em Markov Chain Monte Carlo} simulation \cite{allison2012handling,yuan2010multiple}) used to fill the missing variables, require the known non-missing attributes of the record with missing values to be able to use an inference model. In the case of {\em unknown unknowns}, these assumptions are violated as the entire record (i.e., all attributes) are missing.

Missing data is also well studied in databases \cite{rahm2000data,osborne2012best,lang2014partial}; however, as traditional RDBMS query processing function under the {\em closed world} assumption, they do not consider {\em unknown unknowns} as part of the query processing and largely consider it a data cleaning aspect. 

Recent works \cite{lang2014partial,razniewskiidentifying} defined database completeness in a partly {\em open world} semantic (i.e., database can be incomplete, which causes incorrect query results) and use the completeness information to denote the completeness of query results. Similar in spirit to our work, they investigate the impact on query results of entire database records that may be missing \cite{razniewskiidentifying}; however, they also assume the knowledge of population size (e.g., there are 7 days in a week, there are this many cities in France) to define the completely missing records and measure the completeness.

{\bf Sampling-Based Query Processing:} \label{sec:query_approximation}
To cope with aggregates over large data sets, sampling based estimation techniques have been proposed as part of query processing \cite{olken1986simple,haas1996hoeffding,rice2006mathematical}. One limiting aspect of any sampling based estimation techniques, though, is that they assume a complete database  (i.e., {\em closed world}). 

\section{Conclusion} \label{sec:concl}
Integrating various data sources into a unified data set is one of the most fundamental tools to achieve high quality answers. 
However, even with the best data integration techniques, some relevant data might be missing from the integrated data set.
In this work, we have developed techniques to quantify the impact of any such missing data on simple aggregate query results. The challenge lies in the fact that the existence and the value of the missing data is unknown. 
To our knowledge, this is the first work on estimating the impact of {\em unknown unknowns} on query results.

By nature, our techniques cannot predict black swan events (i.e., extremely rare data items) due to a heavily skewed publicity distribution. 
However, based on our evaluation results, we believe that the proposed techniques can provide valuable insights for users; rather than blindly believing the closed-world query result, the user gets an idea of what the impact of {\em unknown unknowns} might be. 

There are several interesting future directions. 
Currently, none of our estimators provides the best performance under all circumstances. 
The {\em Monte-Carlo} estimator is very robust against streakers, whereas the {\em bucket} estimator provides the most accurate results, if no streakers are present. 
How to develop a robust estimator in all scenarios remains an important area for future work. 
Similarly, developing a tighter upper-bound for aggregate queries would be of great value. 
Finally, extending the proposed techniques for more complex aggregate queries (e.g., with joins) also remains open for future work.
%in this work we focused entirely on simple aggregates. How and if the techniques translate to queries with joins remains an open question. 

This work is an important step towards providing higher quality query results. After all, we live in a big data world where even an integrated data set over multiple sources is possibly incomplete. 

\begin{scriptsize}
\bibliographystyle{abbrv}
\bibliography{main}  
\end{scriptsize}

\appendix

\section{Symbol table} \label{appendix:symbols}
\setlength{\textfloatsep}{1pt}
\begin{table}[h!]
\centering
\small
\begin{tabular}{| l | l |}
\hline
 $\Omega$ & Universe of all valid entities (unknown size)\\
 \hline
 $r$ & A valid unique entity or data item \\
 \hline
 $D$ & Ground truth or the underlying population \\
 \hline
 $S$ & Observed sample of size $n=|S|$, with duplicates \\
 \hline
 $K$ & Integrated database with only unique entities from $S$ \\
 \hline
 $U$ & {\em Unknown unknowns} that exist in $D$, but not in $S$ or $K$ \\
 \hline
 $M_0$ & {\em Unknown unknowns} distribution mass in $D$\\
 \hline
 $c$ & The number of unique data items in $S$; $c=|K|$ \\
 \hline
 $s_j$ & Source $j$ with $n_j=|s_j|$ data items \\
 \hline
 $N$ & The size of the ground truth; $N = |D|$ \\
 \hline
 $\phi$ & The aggregated query result: e.g., $\phi_D$ (over $D$) \\
 \hline
 $\Delta$ & {\em The impact of unknown unknowns}: $\Delta =  \phi_{D} - \phi_{K}$ \\
 \hline
 $f_j$ & A frequency statistic, i.e., the number of data items\\
 & with exactly $j$ occurrences in $S$.\\
 \hline
 $F$ & The set of frequency statistics, $\{f_1,f_2,...,f_n\}$\\
 \hline
 $\rho$ & The correlation between publicity and value distribut-\\ & ions, i.e., {\em publicity-value correlation}\\
 \hline
 $\gamma$ & Coefficient of variance (data skew measure)\\
 \hline
 $C$ & Sample coverage, also $C=1-M_0$ \\
 \hline

\end{tabular}
\caption{Symbols}
\label{tab:abc}
%\vspace*{-10pt}
\end{table}

\vspace{-10pt}

\section{Static Bucket Based Estimator} \label{appendix:static}
In Section~\ref{sec:sum:bucket:static}, we state that the optimal number of buckets depends on the underlying {\em publicity} distribution. Here, we elaborate on this with the two examples.

\begin{figure}[h!]
 \centering
 \vspace*{-10pt}
 \includegraphics[width=2.8in]{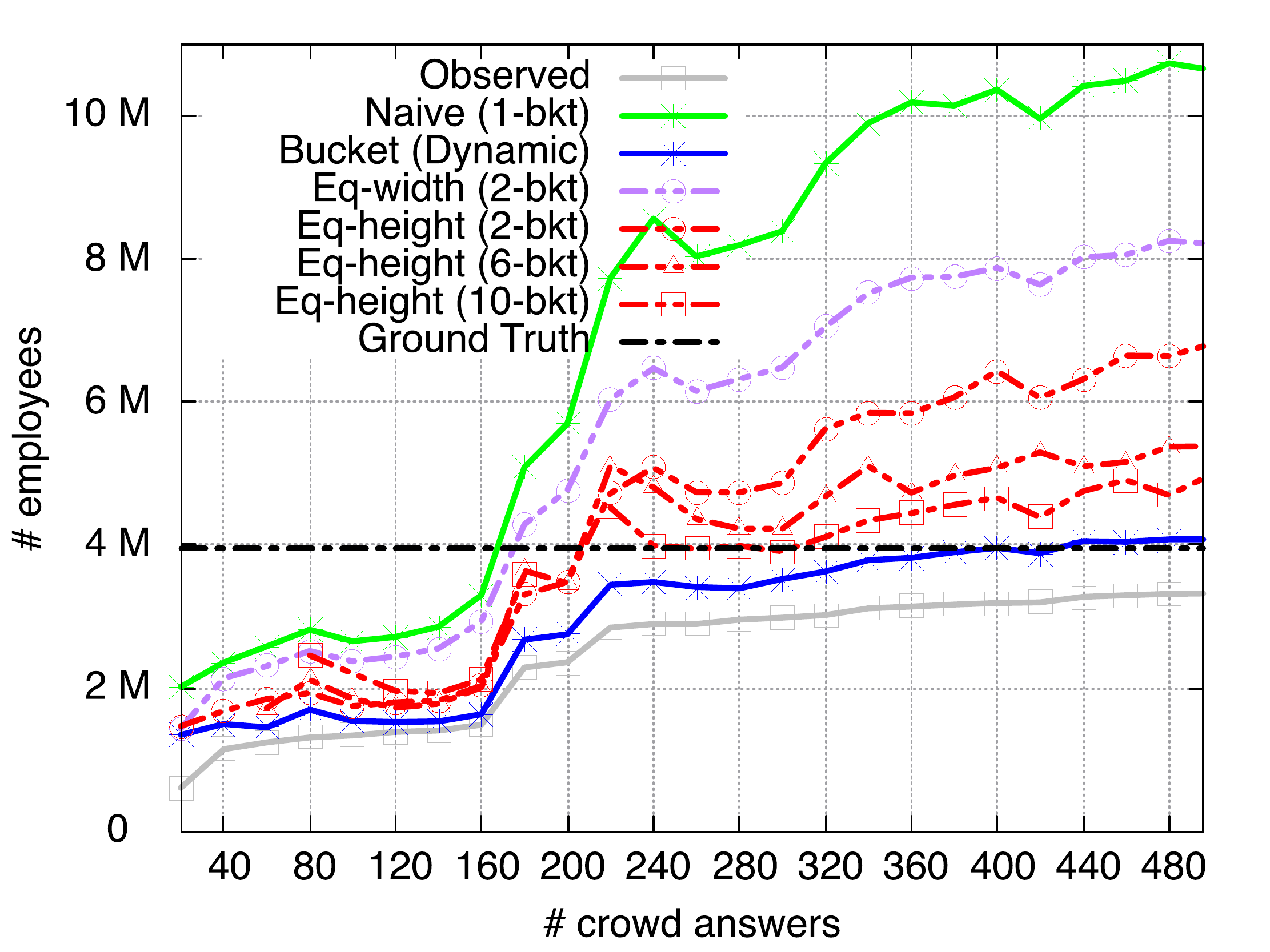}
 \caption{The best US tech-sector employment estimation with static buckets. Splitting into more buckets improves estimation. Eq-width (6-bkt, 10-bkt) are missing due to some of the buckets are empty.}
 \vspace*{-10pt}
 \label{fig:static_buckets}
\end{figure}

%\michael{The individual lines are hard to distinguish. Especially Eq-width and Eq-height. Maybe change the line type for each as well. Also maybe consider ordering all but ground truth in order of placement (makes it easy to determine which is which). These comments apply to the next figure as well.} 

Figure~\ref{fig:static_buckets} shows the US tech-sector employment estimates by various estimators: $Naive$ (1-bucket), $Bucket$ (a.k.a., $Dynamic~Bucket$), and $Static~Bucket$ (Eq-width and Eq-height). In this particular example, splitting into more buckets improves estimation, as the underlying {\em publicity} distribution is skewed and correlated to the values (i.e., larger companies are more well known). 

\begin{figure}[h!]
 \centering
 \includegraphics[width=2.8in]{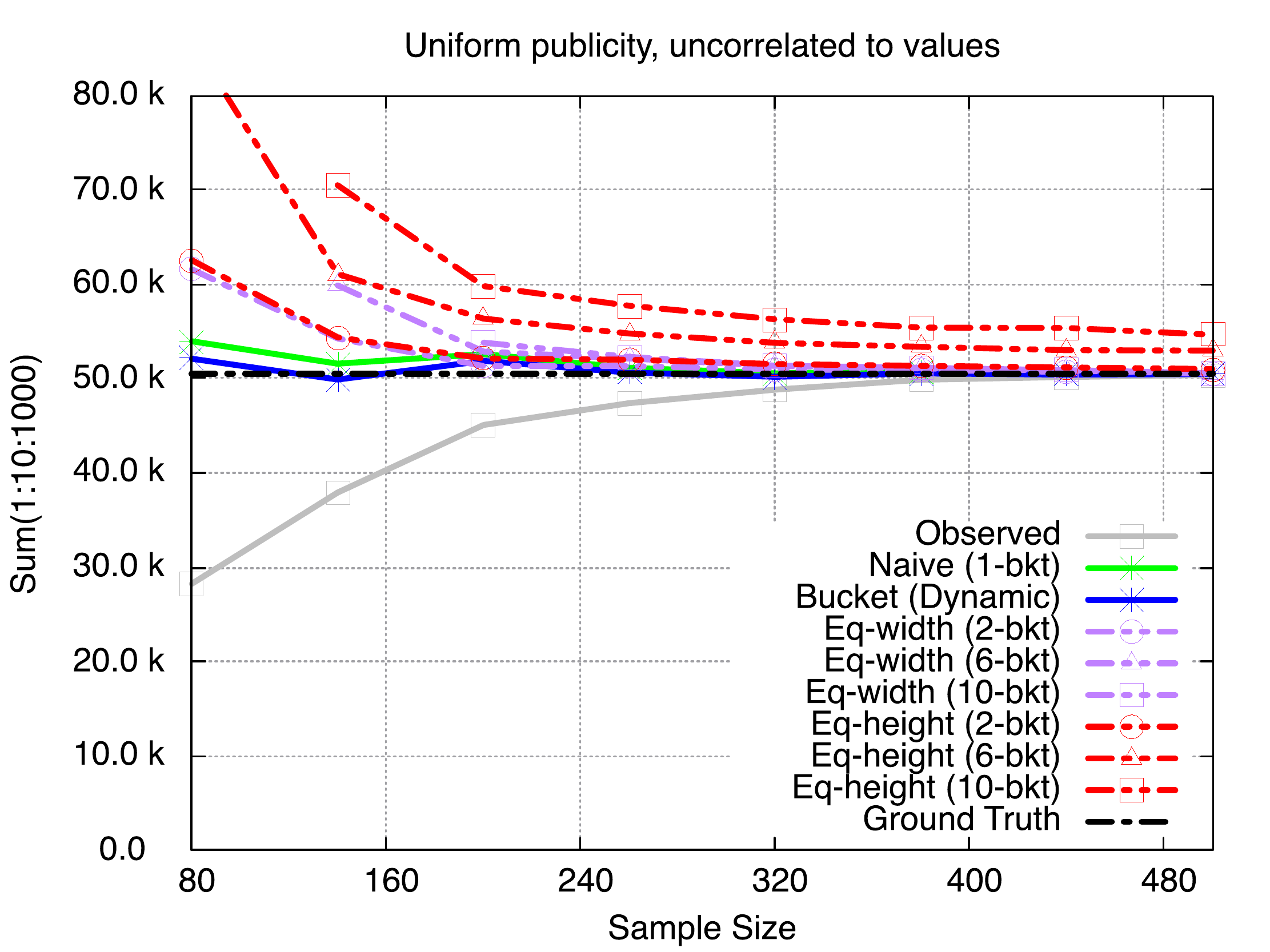}
 \caption{Sum(1:10:1000) estimation with static buckets. Splitting less (e.g., $Naive$) improves estimation. Data points are missing when some buckets contain $singletons$ only (i.e., infinite estimation).} 
 %\vspace*{-18pt}
 \label{fig:static_bucket_unif}
\end{figure}

In contrast, in the simulated case in Figure~\ref{fig:static_bucket_unif}, splitting into less (e.g., $Naive$) improves estimation as the underlying {\em publicity} is uniform. Notice, that in both examples above, the $bucket$ estimator yields the best estimates, dynamically resizing buckets on its own. 
%\michael{Maybe state something like the Bucket estimator splitting less because it determines it to be better. I.e. connect the two sentences a bit more. You're saying splitting less is better and Bucket is best, so explain how Bucket splits less. It's a small thing, but it would read better if done.}

Also notice, that we consider two variants of static buckets: the one described in the paper, {\bf equi-width}, which divides the observed value range into a fixed number of buckets, and another obvious variant, {\bf equi-height}, which divides the observed sample, sorted by value, evenly into a fixed number of buckets. Both static bucket types are simple to use, but they require parameter tuning for the optimal number of buckets, which is hard to predict without knowing the true {\em publicity} distribution. 

\section{The Increase in Count Estimate After Bucket Split} \label{appendix:increase_count}

In equation~\ref{eqn:bucket_ineq_2}, we claimed that the count estimation ($\hat{N}_{Chao92}=nc/(n-f_1)$) of a bucket increases after splitting the bucket, if data items are evenly distributed over the {\em attribute} value range, and there is no {\em publicity-value correlation}:
\begin{equation}
\begin{split}
\hat{N}_{Chao92} & = \frac{c}{1-f_1/n} = \overbrace{\frac{n\cdot c}{n-f_1}}^{\text{Before split}}\\
 & \leq \underbrace{\frac{\frac{n}{2}\cdot\frac{c}{2}}{\frac{n}{2}-\alpha \cdot f_1} + \frac{\frac{n}{2}\cdot\frac{c}{2}}{\frac{n}{2}-(1-\alpha)\cdot f_1}}_{\text{After split}}
\end{split}
\nonumber
\end{equation}

The $\alpha$ parameter governs the split of the original singleton count ($f_1$) into a pair of smaller buckets. We assume $n$ and $c$ are evenly distributed between the split buckets, as items are evenly distributed over the value range, and all values are equally likely (no {\em value-publicity correlation}). We now show that the above inequality holds by showing that the right hand side (after split) is minimized at $nc/(n-f_1)$. Note that $nc/(n-f_1)$ is a positive number as $n\geq f_1 \geq 0$ and $c\geq 0$.

To find the minimum, we take the first derivative of the right hand side (denoted by $\mathcal{R}$) with respect to $\alpha$:

\begin{equation}
\begin{split}
\mathcal{R}' & = \frac{-c\cdot f_1\cdot n}{4(-(1-\alpha)\cdot f_1 + \frac{n}{2})^2} + \frac{-c\cdot f_1\cdot n}{4(-\alpha\cdot f_1 + \frac{n}{2})^2}\\
\end{split}
\nonumber
\end{equation}
Solving $\mathcal{R}'=0$, we get $\alpha=0.5$; we have $\mathcal{R}(0.5)=nc/(n-f_1)$ as shown below:
\begin{equation}
\begin{split}
\mathcal{R}(0.5) & = \frac{\frac{n}{2}\cdot\frac{c}{2}}{\frac{n}{2}-0.5 \cdot f_1} + \frac{\frac{n}{2}\cdot\frac{c}{2}}{\frac{n}{2}-(1-0.5)\cdot f_1}\\
 & = \frac{\frac{n}{2}\cdot\frac{c}{2} + \frac{n}{2}\cdot\frac{c}{2}}{\frac{n}{2}-0.5 \cdot f_1}
 = \frac{n\cdot c}{n-f_1}
\end{split}
\nonumber
\end{equation}

Finally, we show $\mathcal{R}(0.5)=nc/(n-f_1)$ is the minimum by ensuring $\mathcal{R}''(0.5) > 0$:

\begin{equation}
\begin{split}
\mathcal{R}'' & = \frac{c\cdot f_1^2\cdot n}{2(-(1-\alpha)\cdot f_1 + \frac{n}{2})^3} + \frac{c\cdot f_1^2\cdot n}{2(-\alpha\cdot f_1 + \frac{n}{2})^3}\\
\mathcal{R}''(0.5) & = \frac{c\cdot f_1^2\cdot n}{2(-(1-0.5)\cdot f_1 + \frac{n}{2})^3} + \frac{c\cdot f_1^2\cdot n}{2(-0.5\cdot f_1 + \frac{n}{2})^3}\\
 & = \frac{c\cdot f_1^2\cdot n}{(-0.5\cdot f_1 + \frac{n}{2})^3} = \frac{8c\cdot f_1^2\cdot n}{(-f_1 + n)^3}
\end{split}
\nonumber
\end{equation}
Note that $n \geq f_1$, and this makes $\mathcal{R}''>0$; $\mathcal{R}$ is minimized at $nc/(n-f_1)$ and the inequality holds true:

\begin{equation}
\begin{split}
\overbrace{\frac{n\cdot c}{n-f_1}}^{\text{Before split}} & \leq \underbrace{\frac{\frac{n}{2}\cdot\frac{c}{2}}{\frac{n}{2}-\alpha \cdot f_1} + \frac{\frac{n}{2}\cdot\frac{c}{2}}{\frac{n}{2}-(1-\alpha)\cdot f_1}}_{\text{After split}}
\end{split}
\nonumber
\end{equation}

\section{Other Estimators}
\label{appendix:other_estimators}
Many proposed techniques can be combined: we can use the {\em frequency} estimator, instead of the {\em \naive{}} estimator, with the {\em bucket} (i.e., {\em Dynamic Bucket} approach) estimator or the {\em Monte-Carlo} estimator. We can also combine the {\em Monte-Carlo} estimator with the {\em bucket} estimator. 
%\michael{Explain. Why is this more interesting? Just a single sentence or half sentence or something.}

\begin{figure}[h!]
 \centering
 \vspace*{-10pt}
 \includegraphics[width=2.8in]{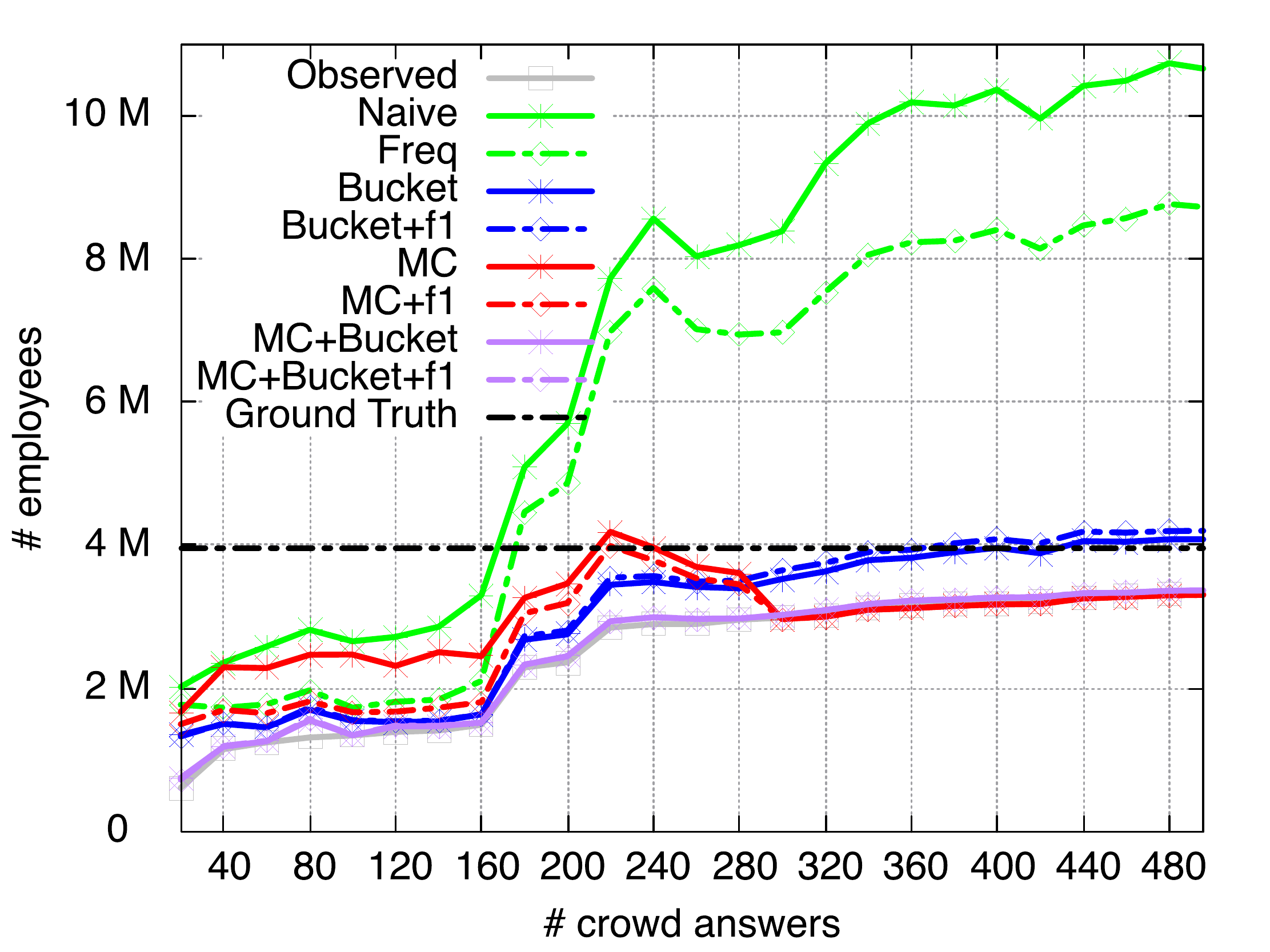}
 \caption{The best US tech-sector employment estimation with other estimators}
 \vspace*{-10pt}
 \label{fig:other_estimators}
\end{figure}

However, as the {\em Monte-Carlo} estimator requires large sample sizes to be accurate, combining it with {\em bucket} estimator often results in lower estimation quality (i.e., each bucket contains a smaller sample). Furthermore, each bucket (a smaller value range) entails a part of the underlying {\em publicity} distribution; hence, the {\em publicity} distribution per bucket appears more uniform. As a major drawback, the {\em Monte-Carlo} estimator exhibits a tendency to favor its count estimate $\hat{N}_{MC} \sim c$ (see Section~\ref{sec:eval:employee}). Such tendency gets more imminent in {\em Monte-Carlo} with {\em Bucket} estimator as seen in Figure~\ref{fig:other_estimators}. Similarly, we found that the difference between the {\em \naive{}}  and {\em frequency} estimators is not significant for the {\em bucket} estimator (i.e., uniform {\em publicity}). 

\section{Number of Sources}
\label{appendix:num_sources}
%{\em Bucket} estimator usually has less assumptions about the underlying distribution but implicitly assumes a single sample without replacement.
{\em Bucket} estimator is non-parametric and works well with with both uniform and skewed distributions; however, it assumes a sample $S$ sampled with replacement.
This assumption is appropriate as long as enough independent data sources contribute evenly to $S$. 

\begin{figure}[!h]
 \subfigure[$w=2$]{\includegraphics[width = 0.5\columnwidth]{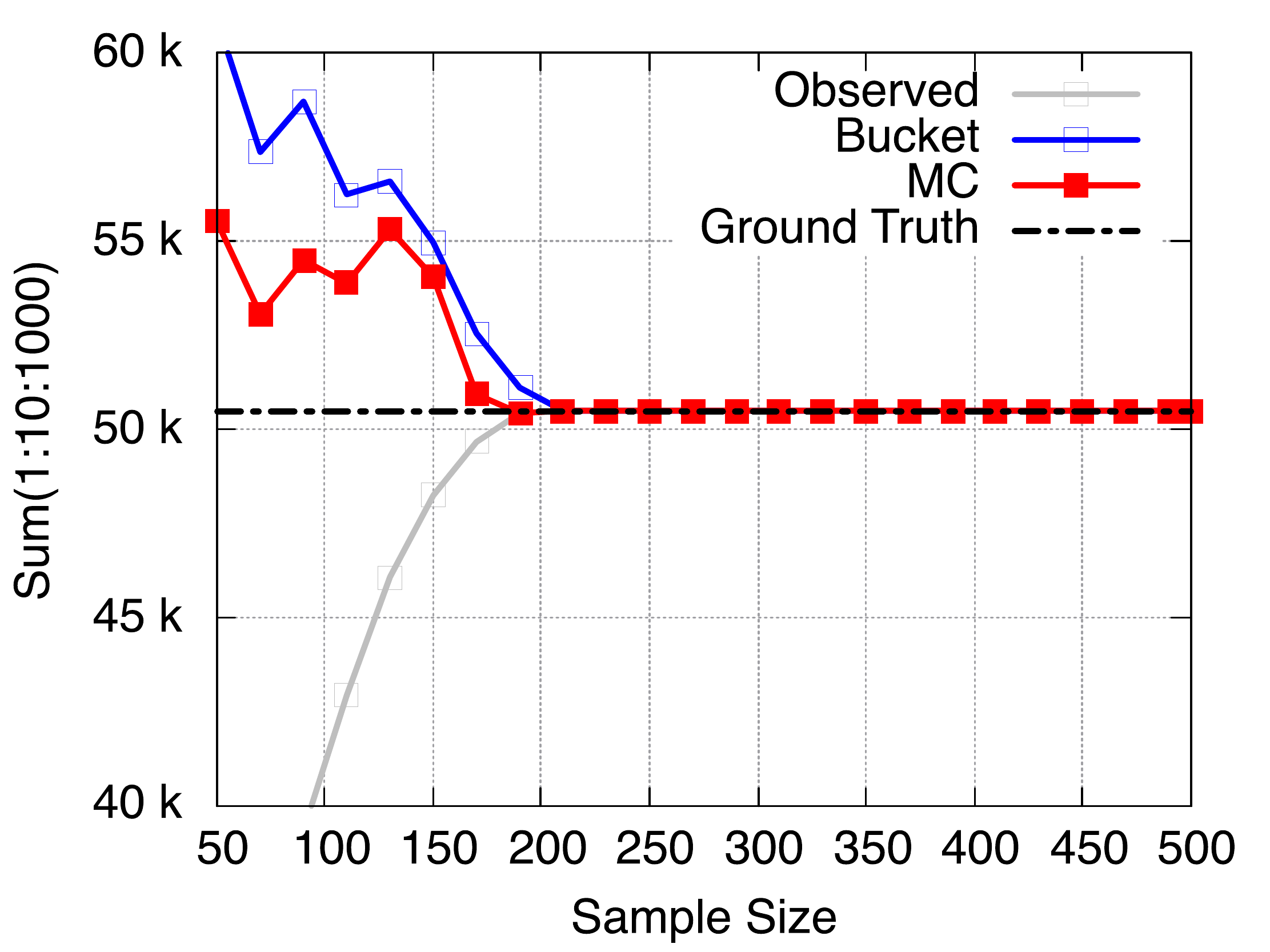}}%
 \subfigure[$w=3$]{\includegraphics[width = 0.5\columnwidth]{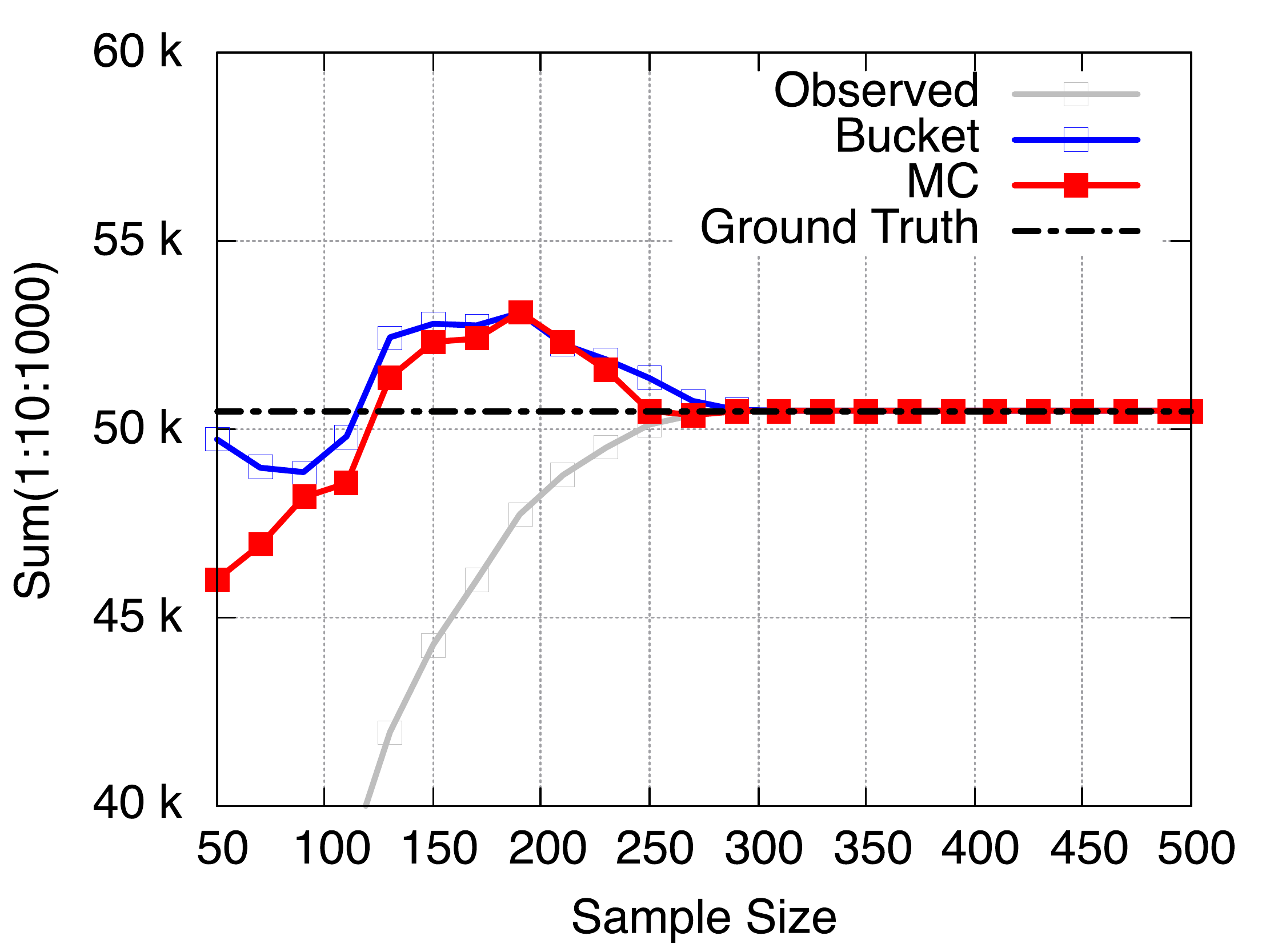}}\vspace*{-10pt}\\
 \subfigure[$w=4$]{\includegraphics[width = 0.5\columnwidth]{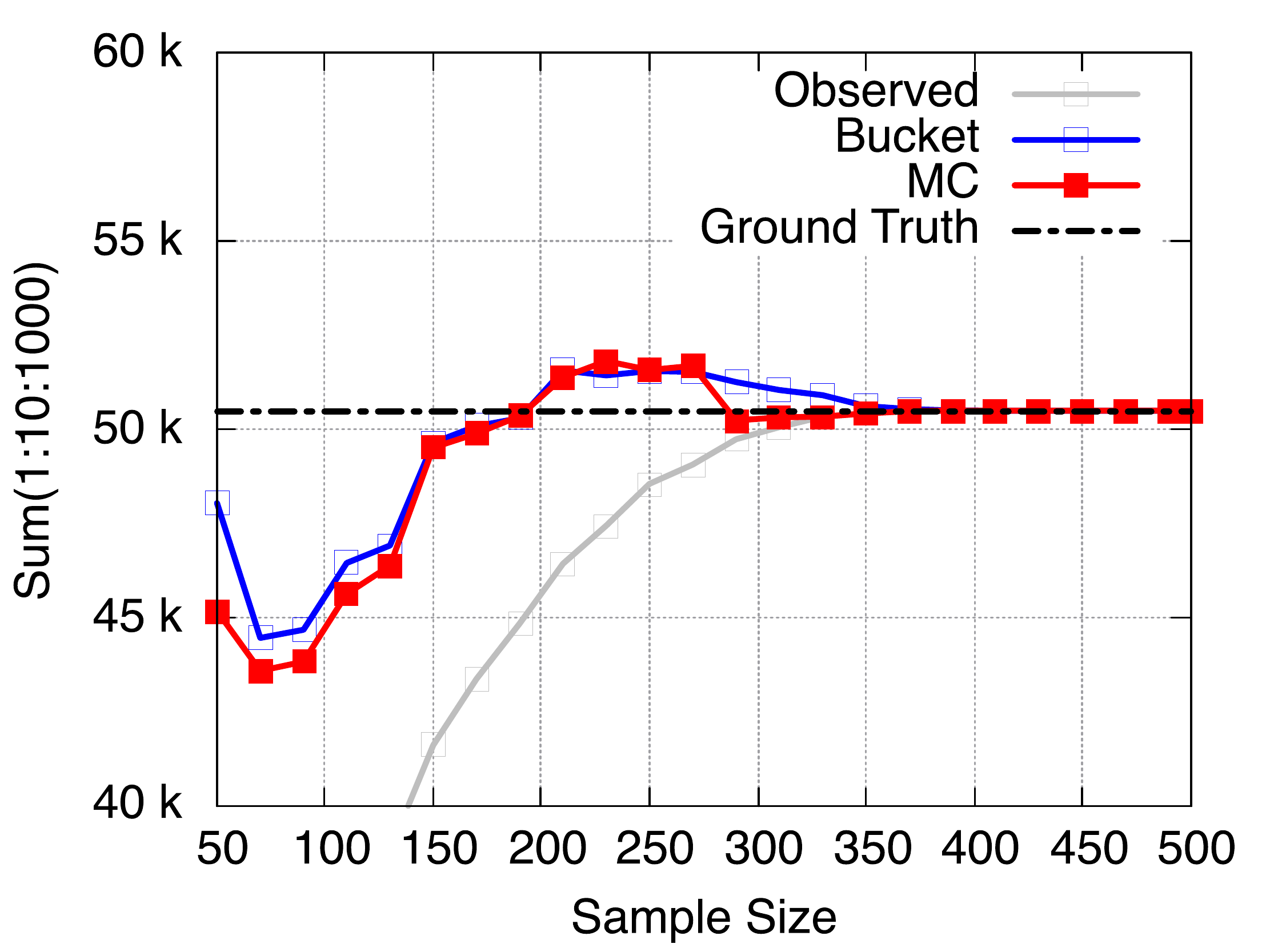}}%
 \subfigure[$w=5$]{\includegraphics[width = 0.5\columnwidth]{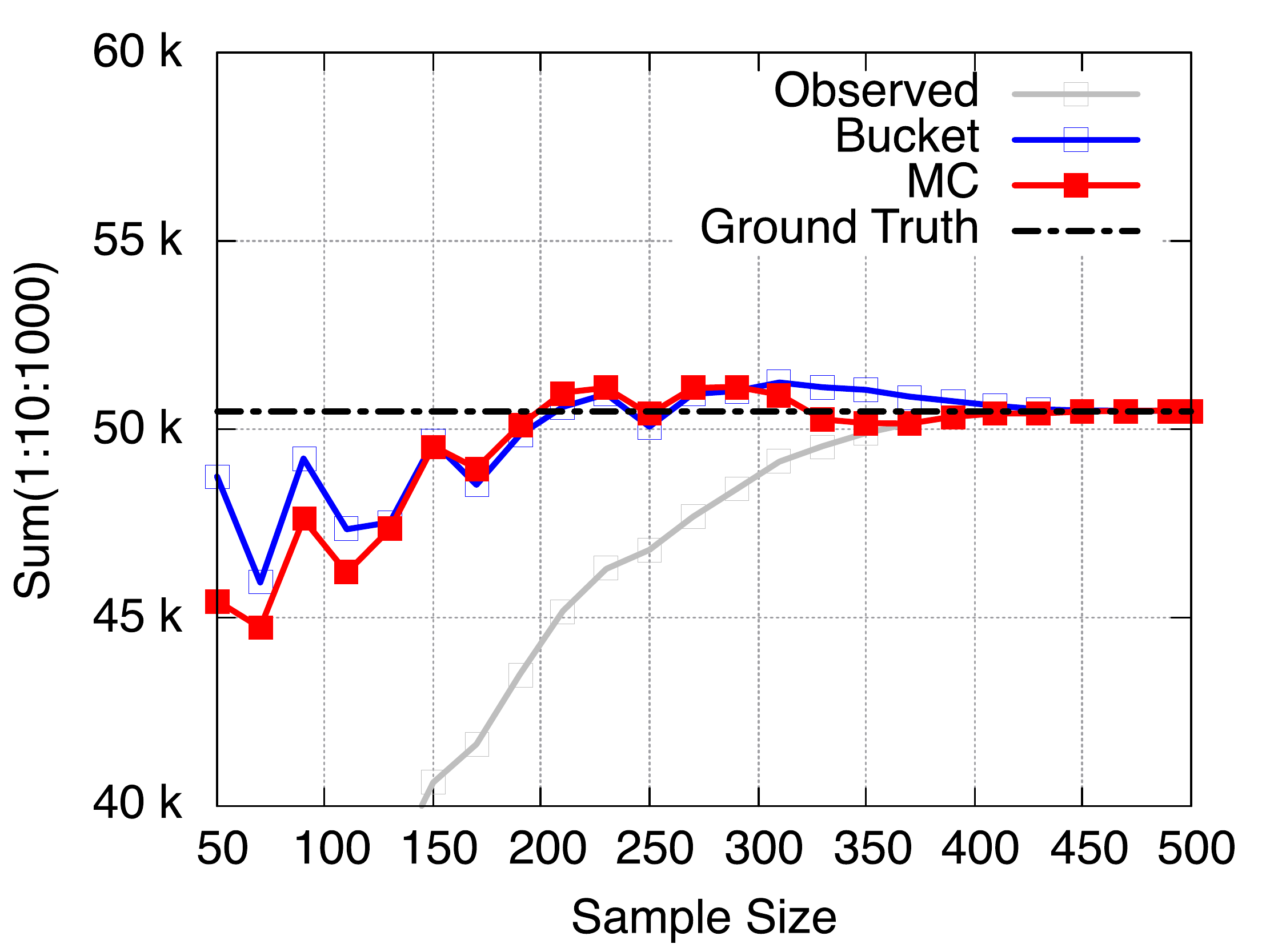}}
 \caption{Synthetic data ($\lambda=4.0$,$\rho=1.0$) with varying number of sources ($w$). {\em Bucket} estimator performs better with more independent sources (i.e., more overlaps).}
 \label{fig:n_src}
\end{figure}

In Figure~\ref{fig:n_src}, we illustrate this with a synthetic data (skewed {\em publicity} correlated to item {\em attribute} values).  
In this particular example, more than 5 sources result in enough overlaps for {\em bucket} to estimate accurately; however, the minimum number of sources would vary with the date set. In addition, {\em Monte-Carlo} estimator converges faster as it does not assume a sample sampled with replacement. 

\section{A Toy Example}
\label{appendix:toy}

\begin{figure*}[!t]
 \subfigure[Multiple sources $s_i$ sampled without replacement from the unknown population $D$. $s_5$ is added later to the original integrated database.]{\includegraphics[width = 0.60\linewidth]{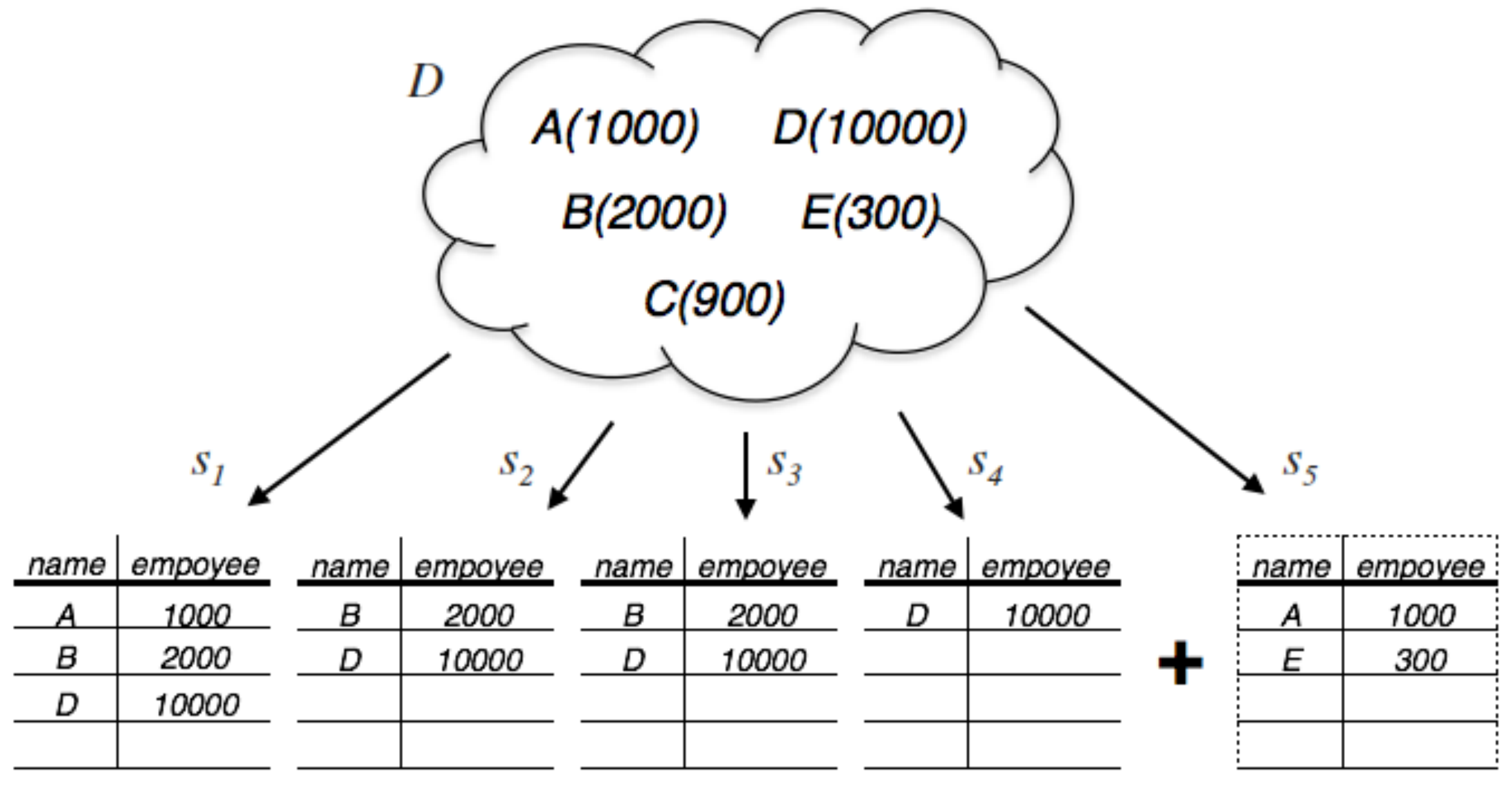}}%
 \hfill
 \subfigure[Integrated Database $K$, before (top) and after (bottom) adding $s_5$]{\includegraphics[width = 0.3\linewidth]{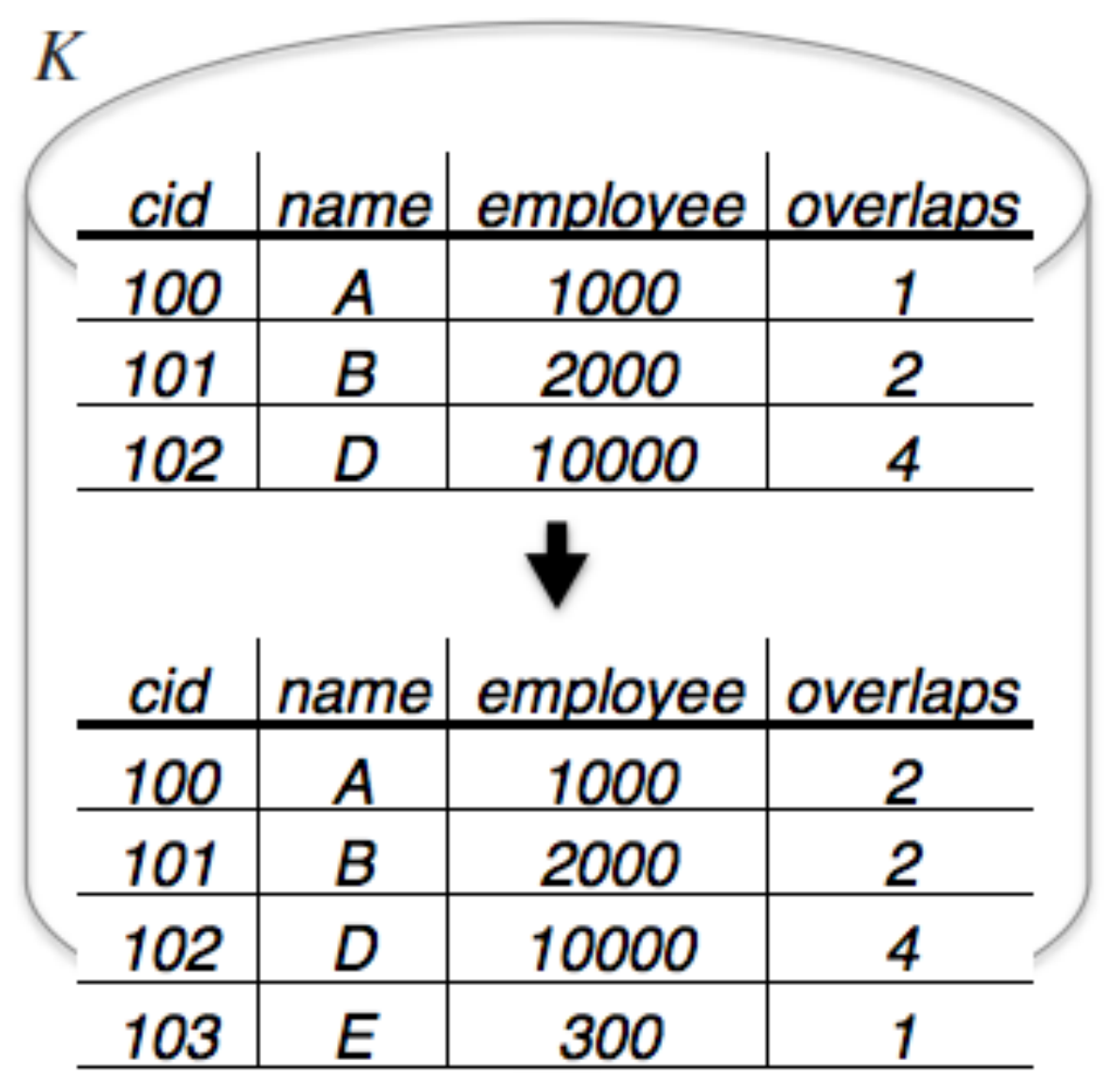}}
   \vspace*{-2pt}
 \caption{A toy example for \texttt{SELECT SUM(employee) FROM K}}
  %\vspace*{-16pt}
 \label{fig:toy_example}
\end{figure*}

\begin{table*}[!t]
\centering
\small
\begin{tabular}{ l | l | l }
 & before adding $s_5$ & after adding $s_5$  \\
 & ($n=7$, $c=3$, $f_1=1$, $\hat{\gamma }^2=0.1667$) & ($n=10$, $c=4$, $f_1=1$, $\hat{\gamma }^2=0$)  \\\Xhline{2\arrayrulewidth}
Ground Truth & \multicolumn{2}{l}{$\phi_D=1000+2000+900+10000+300=14200$} \\\hline
Observed & $\phi_K=1000+2000+10000=13000$ & $1000+2000+10000+300=13300$ \\\hline
Naive & $\begin{aligned}
\phi_K + \Delta_{naive} & = \phi_K + \frac{\phi _K \cdot f_1 \cdot \left(c+\hat{\gamma }^2 n\right)}{c \cdot \left(n-f_1\right)}\\
& = 13000 + \frac{13000 \cdot 1 \cdot \left(3+0.1667\cdot7\right)}{3 \cdot \left(7-1\right)}\\
& \approx 16009
\end{aligned}$ & $\begin{aligned}
& = 13300 + \frac{13300 \cdot 1 \cdot \left(4+0\cdot9\right)}{4 \cdot \left(9-1\right)}\\
& \approx 14962
\end{aligned}$ \\\hline
Freq & $\begin{aligned}
\phi_K + \Delta_{freq} &= \phi_K + \frac{\phi _{f_1} \left(c+\hat{\gamma }^2 n\right)}{n-f_1}\\
& = 13000 + \frac{1000 \left(3+0.1667\cdot 7\right)}{7-1}\\
& \approx 13694
\end{aligned}$ & $\begin{aligned}
& = 13300 + \frac{300 \left(4+0\cdot 9\right)}{9-1}\\
& = 13450
\end{aligned}$ \\\hline
Bucket & $\begin{aligned} \phi_K + \Delta_{bucket} &= \phi_K + \Delta_{b_1:\{A,B\}} + \Delta_{b_2:\{D\}}\\
& = \phi_K + \left\{ \Delta_{naive} \right\}_{b_1} + \left\{ \Delta_{naive} \right\}_{b_2}\\
& = 13000 + \frac{3000 \cdot 1 \cdot \left(2+0\cdot3\right)}{2 \cdot \left(3-1\right)} \\ & + \frac{10000 \cdot 0 \cdot \left(1+0\cdot4\right)}{1 \cdot \left(4-0\right)}\\
& = 14500 \end{aligned}$ &  $\begin{aligned}  &= \phi_K + \Delta_{b_1:\{A,E\}} +\Delta_{b_2:\{B\}} + \Delta_{b_3:\{D\}}\\
& = \phi_K + \left\{ \Delta_{naive} \right\}_{b_1} + \left\{ \Delta_{naive} \right\}_{b_2} + \left\{ \Delta_{naive} \right\}_{b_3}\\
& = 13300 + \frac{1300 \cdot 1 \cdot \left(2+0\cdot3\right)}{2 \cdot \left(3-1\right)} \\ & + \frac{2000 \cdot 0 \cdot \left(1+0\cdot2\right)}{1 \cdot \left(2-0\right)} + \frac{10000 \cdot 0 \cdot \left(1+0\cdot4\right)}{1 \cdot \left(4-0\right)}\\
& = 13950 \end{aligned}$
\end{tabular}
\caption{\texttt{SELECT SUM(employee) FROM K} results with different {\em unknown unknowns} estimators: $bucket$ estimator gives the most accurate estimation of $\phi_D$, } 
\label{tab:toy_result} \vspace*{-15pt}
\end{table*}

In this section, we walk through the different estimators step by step using a simple toy example. 
Again, we use the same query, \texttt{SELECT SUM(employee) FROM K}, from the introduction but over a very simplistic data set, shown in Figure~\ref{fig:toy_example}. 
It should be noted, that this toy example can not convey any statistical properties because of its small size, but we can explain the general reasoning behind the techniques using the example. 

Figure~\ref{fig:toy_example} shows the  data integration scenario of our example.
We assumes that the ground truth $D$ consists of 5 companies  $\{ A,B,C ,D,E \}$ (the bubble on the top), with different numbers of employees (e.g., company $A$ has 1000, whereas company $B$ has 2000). 
In the beginning we have four data sources $\{s_1,s_2,s_3,s_4\}$ each mentioning some of these companies, thus they sample without replacement from $D$.
For instance data source $s_1$ lists companies $A$, $B$, and $D$. 
In the example we also assume a {\em publicity-value correlation}; 
that is, the biggest company $D$ appears in all data sources ($\{s_1,s_2,s_3,s_4\}$), while smaller companies appear in fewer sources. 
To show how the estimates improve, we assume that the data source $s_5$ is added later on (visualized through the plus). 
The tables in Figure~\ref{fig:toy_example}(b) show the integrated database  before (top) and after (bottom) adding the fifth data source.  
For convenience, the last column shows, how many times each company was observed across the multiple data sources. 

Table~\ref{tab:toy_result} shows the estimates by different estimators before and after adding the fifth data source. We exclude {\em Monte-Carlo} estimator due to its simulation based nature. 
The top row contains the relevant statistics of $K$.
For instance, with 4 data sources, the number of observed items / sample size is $n=7$, the number of observed unique items is $c=3$ (i.e., companies $A$, $B$, and $D$ from the top table in Figure~\ref{fig:toy_example}(b)), the number of singletons $f_1=1$ (i.e., company $D$ as it is the only company, which was observed exactly ones across the data sources). and the calculated \emph{coefficient of variance} ($CV$) $\gamma=0.1667$ calculated over the sample.

Before adding the fifth data source, the observed total sum is $\phi_K=1000+2000+10000=13000$, after adding the fifth data source $\phi_K=1000+2000+10000=13300$. 
In this example, the observed total sum does not converge to the ground truth of $14200$

Table~\ref{tab:toy_result} shows the values with calculations for the different estimators. 
As it can be seen, the \naive{} estimator performs the worse; the estimator is quite far off, especially with 4 data sources.
The reason is the value estimator ({\em mean substitution }) used. 
The average number of employees is $\phi_K / 3 \approx 4333$. Thus all missing companies (i.e., {\em unknown unknowns}) are also assumed to be that big. 
Now knowing that bigger companies are more likely to be sampled, now the \naive{} estimator heavily over-estimates.

In contrast, the {\em frequency} estimator performs much better than the \naive{} estimator because it assumes that the missing companies have the average value over {\em singletons}, which includes $A$, but not the extremely big company $D$; the missing companies are assume to have a value of $\phi_{f_1} / 1 = 1000$. Because less popular companies are more likely to be smaller (i.e., the {\em publicity-value correlation}), this yields to a much better estimate. 

Finally, the {\em bucket} estimator performs the best. Before adding the fifth source, the algorithm creates two buckets: $b_1:\{A,B\}$ and $b_2:\{D\}$. The estimate quality of {\em bucket} persists even after we add $s_5$ (i.e., {\em Bucket} is the best). In this case, the {\em bucket} estimator generates $b_1:\{A,E\}$, $b_2:\{B\}$ and $b_3:\{D\}$. 
The {\em bucket} estimator automatically groups the small companies ($A$ and $E$) together and uses their average number of employees for the missing companies (all other buckets have unknown count estimation of $0$); in this example, the {\em bucket} estimator has a smoothed value in between $300$ and $1000$. This is particularly more desirable compared to the case of the {\em frequency} estimator: $E$ is the new one and only {\em singleton} and $\phi_{f_1}$ is now $300$.

\end{document}